\documentclass[journal,10pt]{IEEEtran}

  \usepackage[nocompress]{cite}
  \usepackage{amsmath}
\usepackage{hyperref}
\usepackage{cite}
\usepackage{graphicx}
\usepackage{sidecap}
\usepackage{bigints}
\usepackage{mathrsfs}
\usepackage{booktabs}
\usepackage{caption}
\usepackage{subcaption}
\usepackage{flushend}
\usepackage{enumerate}
\usepackage{xcolor}

\begin{document}
\title{On the Performance Optimization of Two-way Hybrid VLC/RF based IoT System over Cellular Spectrum}

\author{Sutanu Ghosh, \IEEEmembership{Member, IEEE}, and Mohamed-Slim Alouini, \IEEEmembership{Fellow, IEEE}

\thanks{Sutanu Ghosh and M. -S. Alouini are with the King Abdullah University of Science and
Technology, Thuwal, Makkah Province 23955-6900, Saudi Arabia (e-mail:sutanu99@gmail.com;
slim.alouini@kaust.edu.sa).}}
%Copyright (c) 2015 IEEE. Personal use of this material is permitted. However, permission to use this material for any other purposes must be obtained from the IEEE by sending a request to pubs-permissions@ieee.org.  \par

%\markboth{IEEE Transactions on Vehicular Technology,~Vol.~XX, No.~XX, XXX~2020}
%{}

\maketitle

\begin{abstract}
This paper investigates the system outage performance of a useful architecture of two-way hybrid visible light communication/radio frequency (VLC/RF) communication using overlay mode of cooperative cognitive radio network (CCRN). The demand of high data rate application can be fulfilled using VLC link and communication over a wide area of coverage with high reliability can be achieved through RF link. In the proposed architecture, cooperative communication between two licensed user (LU) nodes is accomplished via an aggregation agent (AA). AA can perform like a relay node and in return, it can access the LU spectrum for two-way communications with Internet-of-Things (IoT) device.  First, closed form expressions of outage probability of both LU and IoT communication are established. On the basis of these expressions, optimization problems are formulated to achieve minimum outage probability of both LU and IoT network.
The impacts of both VLC and RF system parameters on these systems outage probability and throughput are finally shown in simulation results.
\end{abstract} 

\begin{IEEEkeywords}
Hybrid VLC/RF systems, cooperative cognitive radio network, outage probability, Internet-of-Things (IoT) network, spectrum sharing.
\end{IEEEkeywords}

\section{Introduction}

The heavy demand of high data rate wireless applications (such as, network conference and telemedicine) and due to numerous device connectivity, growing congestion over radio frequency (RF) spectrum make visible light communication (VLC) an attractive solution to manage the issue of spectrum scarcity and to support high speed information transmission [1]. VLC can provide high bandwidth over a very high frequency band (430-790 THz) for future generation of wireless communication [2]. However due to the limitation of narrow area of coverage and also the challenging issue of uplink transmission over VLC link, the combined utilization of both VLC and RF system is proposed in the existing literature to achieve the potential benefits of high bandwidth, mitigation of the problem of blockage of line-of-sight transmission, provision of wide area of coverage and better uplink transmission of data stream [3]-[4] for
future 6G applications. Additionally, the reliability and significant range of effective communication is also possible to enhance through cooperative communication using intermediate relay nodes [5]-[6]. Those nodes may operate over hybrid VLC/RF system for various indoor/outdoor implementations of wireless communication [7]-[8]. Indoor approach of VLC communication is more popular for light fidelity (Li-Fi) communication [9] and outdoor approach is convenient to implement vehicular network applications [10]. 

The relay nodes are mostly small gadgets like Internet-of-Things devices (IoDs) or sensor nodes which harvest energy either from VLC communication using simultaneous light-wave information and power transfer (SLIPT) protocol [11] or from RF communication using simultaneous wireless information and power transfer (SWIPT) protocol [12] for transmission of their information. SWIPT is typically classified as power splitting (PS), time switching (TS) and hybrid power-time switching (HPTS) schemes [13]. Various studies on one-way relaying using SLIPT protocol is reported in previous works [14]-[16]. In [7], [14]-[16], relay node harvests energy from the received signal over VLC link and forwards the received information to the destination over RF link. The goal of [14] is to maximize the end-to-end data rate of hybrid VLC/RF system and the outage minimization problem is investigated in [7], [16]. In [15], non-orthogonal
multiple-access (NOMA) principle is applied and the strong user of the proposed system performs the relaying action to forward information towards the weak user over RF link.

Furthermore, the issue of energy harvesting (EH) enabled large number of device connectivity in next generation Internet-of-Things (IoT) applications (like,  healthcare, industrial IoT, traffic monitoring, environmental monitoring etc.) is also an important aspect in 5G/6G wireless networks to exchange data between two or more nearest users [17]. Devices may communicate with each other either accessing unlicensed frequency band (like, industrial-scientific-medical band) following out-band approach of device-to-device (D2D) communication or accessing the licensed band of frequency spectrum (like, cellular spectrum) following in-band approach of D2D communication [18]-[19].

Moreover, the efficient usage of frequency spectrum is desirable to enhance the spectrum efficiency (SE) of wireless networks [20]. SE is possible to increase by implementing two-way communications over one-way communication. In addition to that, the utilization of two-way communications applying the spectrum sharing strategies (interweave, underlay and overlay) of cognitive radio (CR) technology is also beneficial to achieve high SE [21].
To the best of our knowledge, the performance of two-way communications in cooperative CR network (CCRN) over VLC/RF system has not been investigated earlier. Motivated by this issue, the current work studies the system  performance of the communication between aggregation agent (AA) and IoD over licensed spectrum using CR-overlay mode of spectrum sharing and hybrid VLC/RF EH technique. The architecture is analysed for outdoor VLC applications.

{\bf{Scope and Contributions:}}
{{In the existing literature on EH, the research works consider EH either from VLC link or from RF link. However, the combined usage of these two different sources is not considered together for EH. In addition to that, previously all the existing works related to hybrid VLC/RF communication consider one-way communication. This work, on the contrary, explores the issue of two-way cooperative communications between VLC system and platform server (PSe) and two-way IoD communications over licensed spectrum using SLIPT and SWIPT enabled DF relaying mechanism to enhance SE of the system. In case of RF-EH, HPTSR protocol is followed to utilize both time and power in an effective manner to maximize both throughput and reliability of two-way licensed user (LU) communications and two-way IoT communications. The main contributions of this work can be summarized below.

$\bullet$  A novel four-phase two-way LU communication and two-way IoT communications using harvesting energy from different energy sources (from VLC system over VLC link and from PSe over RF link) is proposed to enhance the utilization of VLC and radio resources. %In RF EH, HPTSR protocol is used to achieve two fold benefits of both PS and TS approaches.

$\bullet$ Overlay mode of CR framework is used by AA and IoD for sharing the spectrum of LU network. Closed form analytical outage expressions of both LU communications and IoT communications are derived.

$\bullet$ Optimization framework is also developed to maximize throughput of both LU communications and IoT communications and to minimize the outage of both LU and IoT communications to boost up reliability of the proposed system. The optimal range of time allocation for EH from both VLC link and RF link is achieved to realize the effectiveness of two different links in the hybrid system.

$\bullet$ Finally, numerical results provide the usefulness of various parameters such as PS factor, TS factor, LED power, RF power on the performance of the proposed system.

Various symbols and their meaning are shown in Table-I. The rest of the portion of this paper is arranged as follows. The system model and the operational procedure are discussed in Section II. Both licensed network and IoT network outage expressions are derived and optimization problem of the minimization of both the outage probabilities is also formulated in Section III. The numerical results are shown in Section IV, and finally the paper is concluded in Section V.}}
 
\begin{table}\caption{Symbols and Meaning}
  \centering
    \begin{tabular}{l|p{63mm}}
    \hline
    \hline
    Symbols & Definitions\\
    \hline
    \hline
    $N_L$ & Number of LEDs present in LED array\\
    \hline
    $P_{L}$ & LED power\\
    \hline
    $B_{dc}$ & DC bias current\\
    \hline 
    $A_p$ & Peak amplitude of transmitted signal of VLC system\\
    \hline
    $\vartheta$ & Optical to electrical conversion efficiency at AA\\
    \hline
    $h_{vc}$ & VLC channel co-efficient\\
    \hline 
    $h_L$ & Path loss of VLC channel\\
    \hline 
    $\theta_{FOV}$ & Angle of field-of-view\\
    \hline
    $A_s$ & Physical area of photo-detector\\
    \hline
    $\varsigma$ & Lambert index\\
    \hline   
    $\Theta$ & Irradiance angle\\
    \hline
    $\Phi$ & Incidence angle\\ 
    \hline  
    {{$d_t$}} & {{Distance between VLC system and AA}}\\ 
    \hline
    $f$ & Fill factor\\
    \hline
    $\tau$ & Time allocation for EH over VLC channel\\
    \hline
    $\rho$  &  PS factor of AA\\
    \hline
    $\nu$  &  Path loss exponent of RF link\\
    \hline
    $\eta$ & RF-to-DC conversion efficiency at AA\\
    \hline
    $\varrho$ & Power allocation factor at AA\\
    \hline
    $P_m$  & Maximum transmit power of PSe\\
    \hline
    $d_m$  & Distance between PSe and AA\\
    \hline
    $h_i$ & Channel co-efficient between AA and IoD\\
    \hline
    $h_r$ & Channel co-efficient between PSe and AA\\
    \hline
    $P_{bt}$ & Transmitted power from power transmitter\\
    \hline
    ${h}_{bt}$ & Channel co-efficient between power transmitter and IoD\\
    \hline
    ${d}_{bt}$ & Distance between power transmitter and IoD\\
    \hline
    \hline
    \end{tabular}
  \end{table}
%\begin{figure}[t]
   %  \centering
    % \begin{subfigure}[b]{0.485\textwidth}
      %   \centering
        % \includegraphics[width=\textwidth]{System_model_VLC_RF}
    %    % \caption{\textbf{Outage probability vs $\lambda_1$=$\lambda_2$=$\lambda$ for different values of N$_{tb}$ and m$_k$=1 using hybrid relaying,}}
    %      \end{subfigure}
    % \hfill
     %\begin{subfigure}[b]{0.38\textwidth}
      %   \centering
    %     \includegraphics[width=\textwidth]{APPLI_B}
   %      %\caption{\textbf{for different relaying techniques using N$_{tb}$=m$_k$=2}}
         %     \end{subfigure}
       %         \hfill
       %            \caption{\text{(a) System model, (b) IoT application}}
  %   \hfill
 %   \end{figure}
 %\begin{figure}[h]
 % \centering
  %\includegraphics[width=01\linewidth]{}
  %\caption{\textbf{System model}}
 %\end{figure} 
 
% \begin{figure}[h]
%  \centering
%  \includegraphics[width=01\linewidth]{System_model_rev_1_ck.eps}
 % \caption{\text{{{System model}}}}
% \end{figure} 
 \begin{figure}[h]
     %\centering
     \begin{subfigure}[t]{0.485\textwidth}
         \centering
         \includegraphics[width=\textwidth]{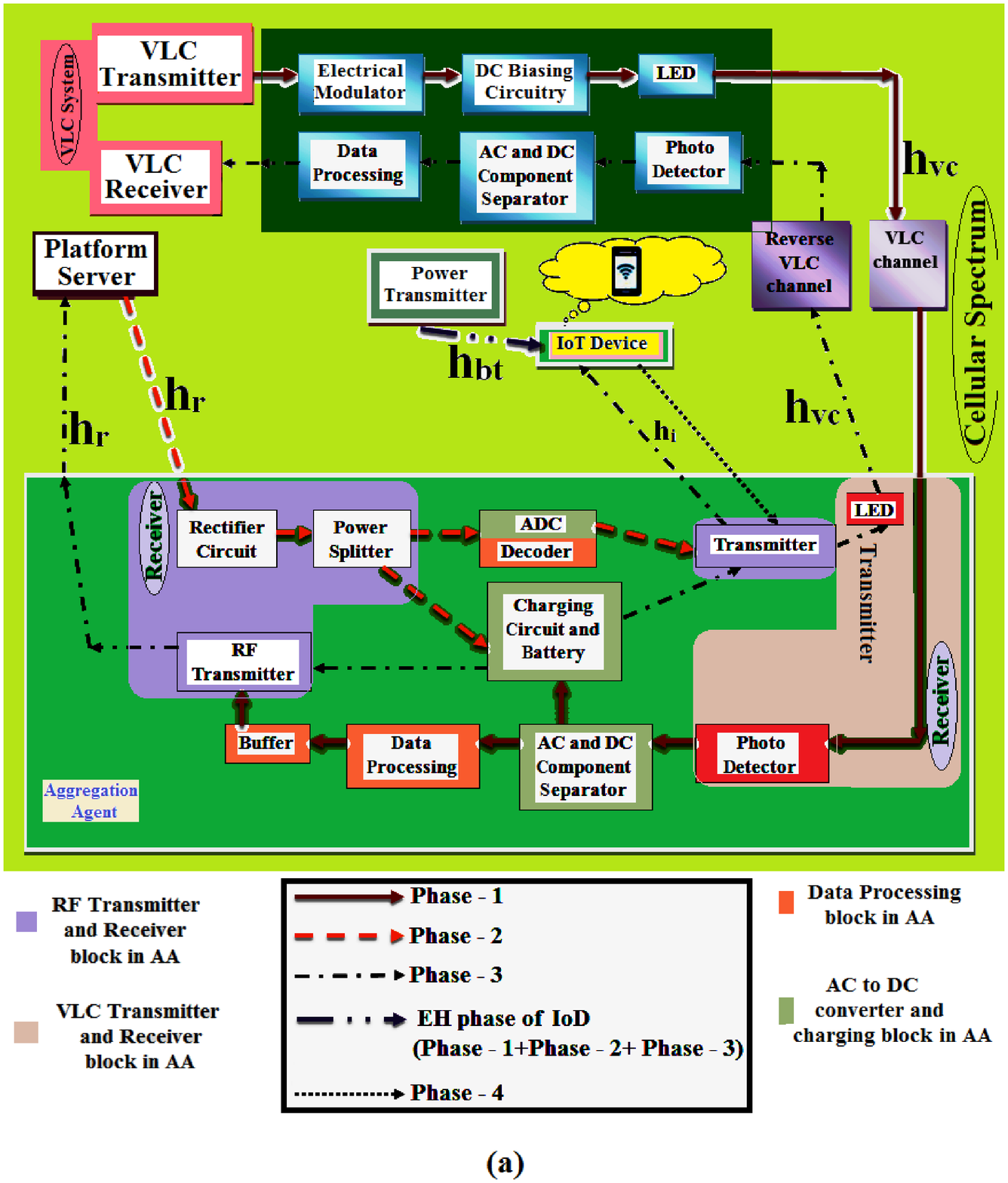}
        % \caption{\textbf{Outage probability vs $\lambda_1$=$\lambda_2$=$\lambda$ for different values of N$_{tb}$ and m$_k$=1 using hybrid relaying,}}
          \end{subfigure}
         \begin{subfigure}[t]{0.49\textwidth}
         \centering
         \includegraphics[width=\textwidth]{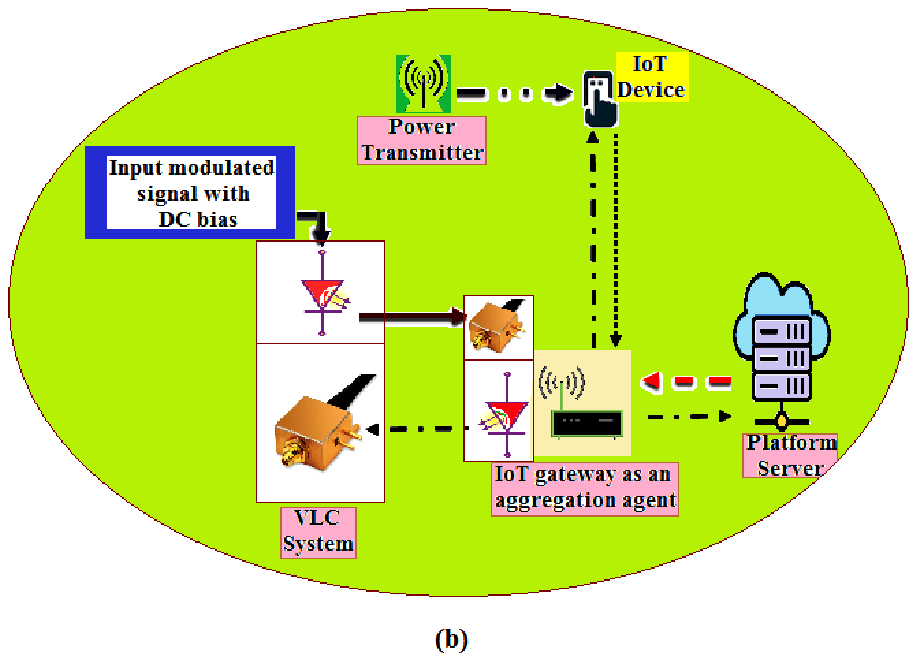}
         %\caption{\textbf{for different relaying techniques using N$_{tb}$=m$_k$=2}}
              \end{subfigure}
                              \caption{{(a) Internal configuration with channel description of system model, {{(b) Simplified form of system model}}}}
        \end{figure}  
 \begin{figure}[h]
  \centering
  \includegraphics[width=0.85\linewidth]{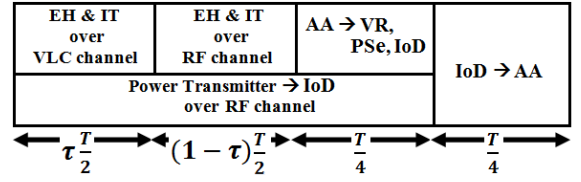}
  \caption{\text{Time frame structure}}
 \end{figure} 
 
 \section{System Model and Operational Procedure}
 Fig. 1.a and Fig. 1.b show two-way communications between a pair of LUs (VLC system and PSe) and co-existing unlicensed user nodes (AA and IoD). The system can be modelled following the overlay mode of CCRN [22]. 
 \begin{enumerate}
 \item VLC transceiver system consists of one VLC transmitter (VT) and one VLC receiver (VR) [23].
The VT has one light emitting diode (LED) array. A VLC link is considered between VLC system and AA and an RF link is considered between PSe and AA [24]. 

\item Due to small coverage area of the VLC link and poor as well as heterogeneous channel condition over the RF link between VLC system and PSe, the cooperation from neighbour node (AA) is needed. In return of this cooperation, AA can access the licensed cellular spectrum to transmit and to receive information from IoD [25]. An RF link is considered between AA and IoD. 

\item AA and IoD are considered to be energy constrained nodes. AA harvests energy from the signal transmission of both VT and PSe over VLC link using SLIPT protocol and over RF link using SWIPT protocol, respectively. IoD harvests energy from power transmitter using wireless power transfer (WPT) protocol. 

\item Time frame structure of the proposed system is shown in Fig. 2. In the first phase, VT transmits signal to AA and in the second phase, PSe sends its signal to AA. 

\item Transmission from VLC system and PSe to AA are considered in two different phases to avoid the problem of decoding at AA. 

\item AA re-encodes (Ex-OR operation is performed for encoding purpose) the decoded received information from the LU nodes. It transmits combined data to both the LU nodes and also sends its own operational data to IoD in a third phase. 

\item The IoD transmits its information to AA in a fourth and last phase. 

\item During the first two phases, AA harvests energy. During the first three phases, the IoD harvests energy. 
\end{enumerate}

The signal descriptions are provided in the subsequent portion of this section.

Signal transmission from VT LED source can be expressed as [14], [26]
\begin{align} 
Y_{S_{T}}^{(1)}(t)=\sqrt{N_L P_{L}} \{s(t)+B_{dc}\},
\end{align}
where $N_L$ is number of LEDs present in LED array system, and $P_{L}$ is the power of each LED considering identical structure of LEDs. $s(t)$ is the RF subcarrier to carry binary data. In (1), $B_{dc}$ is the direct current (DC) bias current and $I_l$, $I_h$ are the minimum and the maximum input bias current.

After the conversion from optical to electrical form, received electrical signal at photo-detector of AA can be expressed as 
\begin{align} 
Y_{PD_{R}}^{(1)}(t) = \underbrace{\vartheta h_{vc}{h_L}[\sqrt{N_L P_{L}} \{s(t)+B_{dc}\}]}_{i_s(t)+I_{D}} + n_{vc}(t), 
\end{align}
where $\vartheta$ indicates optical to electrical conversion efficiency of the receiver at AA, $h_{vc}$ is the VLC channel co-efficient. The path loss of VLC channel can be modelled as $h_L=\frac{A_s(\varsigma+1) \cos^{\varsigma}(\Theta)\cos(\Phi)}{2\pi d_t^2}$ (for, $0 \leq \Phi \leq \theta_{FOV}$) and $h_L=0$ (for, $\Phi \geq \theta_{FOV}$) [27]. 

$\theta_{FOV}$ indicates the angle of field-of-view (FOV). The symbol $A_s$ indicates physical area of photo-detector, $\varsigma$ is Lambert index which depends on semi-angle at half luminance $\Theta_{1/2}$ and $\varsigma=-\frac{1}{\log_2^{cos(\Theta_{1/2})}}$, $\Theta$ signifies irradiance angle, $\Phi$ is the incidence angle. $d_t$ is the distance travelled from VLC system to AA. 
$n_{vc}(t)$ is additive white gaussian noise (AWGN) at AA.

 $I_{D}$ is DC component of $Y_{PD_{R}}^{(1)}(t)$ and it can be defined as $I_{D}=\vartheta h_{vc}{h_L}\sqrt{N_L P_{L}}B_{dc}$. $i_s(t)$ is the alternating current (AC) component of $Y_{PD_{R}}^{(1)}(t)$ and it can be defined as $i_s(t)=\vartheta h_{vc}{h_L}\sqrt{N_L P_{L}}s(t)$. AWGN variance is {{$\sigma_1^2 \approx N_s W=qI_nW$}}, where $q$ is the electronic charge ($1.6 \times 10^{-19}$ c), $I_n$ the induced current in photo-detector come from ambient light source and $W$ is the effective noise bandwidth. 

The received electrical signal power at photo-detector $P_d^e=(\vartheta h_{vc}{h_L}\sqrt{N_L P_{L}} A_p)^2$, where $A_p$ is the peak amplitude of $s(t)$. $A_p=B_{dc}-I_l$, for, $B_{dc}<\frac{I_h+I_l}{2}$ and $A_p=I_h-B_{dc}$, for, $B_{dc}\geq\frac{I_h+I_l}{2}$. {{The instantaneous signal-to-noise ratio (SNR) at photo-detector can then be expressed as 
\begin{align}
\gamma_p=\frac{P_d^e}{\sigma_1^2}=\frac{(\vartheta \mid h_{vc} \mid {h_L} \sqrt{N_L P_{L}} A_p)^2}{qI_nW}.
\end{align}}}

The maximum available harvested energy (in first phase) at AA is $E_{h1}=f I_D V_o \frac{\tau T}{2}$, where $V_o=V_T ln \big( \frac{I_D}{I_o}+1 \big)$. $V_T $ is thermal voltage and $V_T \approx 25 mV$. $I_o$ is dark saturation current of photo-detector and it may vary between 10$^{-9}$ to 10$^{-12}$. $f$ is the fill factor and its value is to be considered for the subsequent section as 0.75. 

For simplicity, the upper bound of $E_{h1}$ is considered as $f V_T  \frac{I_D^2 \tau T}{2 I_o}$ and the harvested power in the first phase is used for data transmission in third phase. The maximum harvested power in first phase can be defined as $P_{h1}=\frac{E_{h1}}{T/4}$.

The received signal in second phase at AA from PSe over RF channel can be expressed as 
\begin{align}
Y_{I_{R}}^{(2)}(t)=\sqrt{\frac{P_{m}}{(d_{m})^{\nu}}} {h}_{r}  X_{m}(t) + n_{r}(t).
\end{align}
$X_{m}(t)$ is the transmitted signal from PSe to AA. AA can harvest the energy from RF signal following HPTS scheme of SWIPT protocol and PS factor is defined by $\rho$ (0 $< \rho <$ 1). The maximum available harvested energy (in second phase) at AA can be expressed as $E_{h2}=\eta \rho \frac{P_{m}}{(d_{m})^{\nu}} \mid{h}_{r}\mid^2 \frac{(1-\tau) T}{2}$. The symbol $\eta$ (0 $< \eta \leq$ 1) indicates RF-to-DC conversion efficiency. $P_m$ is the maximum transmission power of PSe. $d_m$ is the distance between PSe and AA and $n_{r}(t)$ is the channel noise over RF link. The maximum harvested power in second phase can be defined as $P_{h2}=\frac{E_{h2}}{T/4}$. RF link path loss exponent is defined by $\nu$. 

Rest $(1-\rho)$ fraction of received power from PSe is used for decoding received information in second phase. The received signal for information decoding is written as
\begin{align}
Y_{I_{R}}^{(2I)}(t)=\sqrt{\frac{(1-\rho)P_{m}}{(d_{m})^{\nu}}} {h}_{r}  X_{m}(t) + n_{r}(t).
\end{align}
{{The instantaneous SNR at AA from PSe is $\gamma_2=\frac{(1-\rho)P_{m} \mid {h}_{r} \mid^2}{(d_{m})^{\nu} \sigma_2^2 }$}} and {{$\sigma_2^2$ is noise variance over RF channel. Total harvested power in first (using SLIPT protocol over VLC link) and second (using SWIPT protocol over RF link) phase can be combined and expressed as $P_h=P_{h1}+P_{h2}$.}}

{{In the third phase, AA uses the harvested power and divide it into two parts based on the power allocation factor $\varrho$ (0 $<$ $\varrho$ $<$ 1). $\varrho$ portion of $P_h$ is applied for relaying network-coded LU information and remaining $(1-\varrho)$ portion is applied for transmitting its own information $X_{I_{1}}(t)$ to IoD. The combined signal of AA can be expressed as 
\begin{align} 
X_{c_3}(t) = \sqrt{\varrho P_{h}}\{s(t) \oplus X_{m}(t)\} + \sqrt{(1-\varrho)P_{h}} X_{I_{1}}(t).
\end{align}}}

Now, transmitted signal from LED transmitter of AA to VR of VLC system is expressed as 
\begin{align} 
Y_{I_{T}}^{(3)}(t)=\sqrt{\varrho P_h  } \{w(t)+B_{dc}\} + \sqrt{(1-\varrho)P_{h} } \{X_{I_{1}}(t)+B_{dc}\},
\end{align}
where $w(t)$ is the network coded signal and it can be defined as $w(t)= s(t) \oplus X_{m}(t)$.

As self-interference components are known at both VLC system and PSe. Therefore, it is possible to remove by applying self-interference cancellation (SIC) technique. 

After optical to electrical conversion, the received signal at photo-detector of VR can be expressed as 
\begin{multline} 
\hspace*{-0.38cm}Y_{B_{R}}^{(3)}(t) = \underbrace{\vartheta h_{rvc}h_L[\sqrt{\varrho P_h  }\{w(t)+B_{dc}\}]}_{\text{Desired}} \\+ \underbrace{\vartheta h_{rvc} h_L[\sqrt{(1-\varrho)P_{h} } \{X_{I_{1}}(t)+B_{dc}\}]}_{\text{Interference}} + \underbrace{n_{vc}(t)}_{\text{Noise}},
\end{multline}
where $h_{rvc}$ reverse VLC channel co-efficient. The received electrical signal is used for information processing at VR. Now signal-to-interference and noise ratio (SINR) at photo-detector of VR can be expressed as 
%After ignoring the DC part of AC and DC separator at BS, all t
{{\begin{align}
\gamma_3=\frac{\vartheta^2 \mid h_{rvc} \mid^2 h_L^2\varrho P_h  A_p^2}{\vartheta^2 \mid h_{rvc} \mid^2 h_L^2(1-\varrho) P_h  A_p^2+\sigma_1^2}.
\end{align}}}

The received signal at PSe can be written as 
\begin{multline} 
Y_{m_R}^{(3)}(t) = \underbrace{\sqrt{\frac{\varrho P_{h}}{(d_{m})^\nu}}{h}_{r} w(t)}_{\text{Desired}} +\underbrace{\sqrt{\frac{(1-\varrho) P_{h}}{(d_{m})^\nu}} {h}_{r} X_{I_{1}}(t)}_{\text{Interference}}\\+ \underbrace{n_{r}(t)}_{\text{Noise}}.
\end{multline}

{{Now, the instantaneous SINR at PSe can be expressed as 
\begin{align}
\gamma_4=\frac{\varrho  P_h \frac{\mid h_{r} \mid^2}{(d_{m})^\nu}}{ (1-\varrho) P_h \frac{\mid h_{r} \mid^2}{(d_{m})^\nu} +\sigma_2^2}.
\end{align}}}

{{More harvested power (using SLIPT protocol in Phase-1 and SWIPT protocol in Phase-2) is allocated for the transmission of LU information and less harvested power is allocated for IoT information transmission from AA by setting $\varrho>0.5$.}} 
Received signal at IoD in third phase can then be written as
\begin{multline} 
Y_{I_2}^{(3)}(t) = \underbrace{\sqrt{\frac{(1-\varrho)P_{h}}{d_i^\nu}}{h}_{i} X_{I_{1}}(t)}_{\text{Weak signal}} +\underbrace{\sqrt{\frac{\varrho P_{h}}{d_i^\nu}} {h}_{i}  w(t)}_{\text{Strong signal}}+ \underbrace{n_{i}(t)}_{\text{Noise}}.
\end{multline}

%After receiving two signals, IoD$_2$ can remove the strong signal by perfect decoding of received cellular signal and the device can also separate the desired IoT information (received weak signal at smart TV receiver) from strong cellular signal. 
% is considered at IoD$_2$ (because it is able to overhear in previous two phases) and IoD$_2$ can separate desired signal from cellular interference. 
Based on the power level of two received signals, IoD can separate the stronger signal by perfect decoding of received LU signal and the desired IoT information (received weak signal). 
 Finally, it can remove the stronger signal and therefore, received signal at IoD is re-expressed as 
\begin{align} 
Y_{I_2}^{(3)}(t) = \underbrace{\sqrt{\frac{(1-\varrho)P_{h}}{d_i^\nu}}{h}_{i} X_{I_{1}}(t)}_{\text{Desired }}+ \underbrace{n_{i}(t)}_{\text{Noise}}.
\end{align}

{{The instantaneous SNR at IoD from AA is expressed as
\begin{align}
\gamma_5=\frac{(1-\varrho)  P_h \frac{\mid h_{i} \mid^2}{(d_{i})^\nu}}{ \sigma_i^2},
\end{align}
where $h_i$, $n_i(t)$ and $\sigma_i$ indicate channel co-efficient between AA and IoD, channel noise and noise variance, respectively. }}

Now, the harvesting energy at IoD from power transmitter (following WPT principle) can be written as 
\begin{align}
E_{h3}=\eta  \frac{P_{bt}}{(d_{bt})^{\nu}}  \frac{3T}{4} \mid{h}_{bt}\mid^2,
\end{align} 
where $P_{bt}$, ${h}_{bt}$ and ${d}_{bt}$ indicate transmitted power from power transmitter, channel co-efficient of the link between power transmitter and IoD and the distance between power transmitter and IoD, respectively.

The harvested power at IoD ($P_{h3}=\frac{E_{h3}}{T/4}$) is applied in fourth phase for signal transmission from IoD to AA. The received signal at AA can be written as 
\begin{align} 
Y_{I_1}^{(4)}(t) = \underbrace{\sqrt{\frac{P_{h3}}{d_i^\nu}}{h}_{i} X_{I_2}(t)}_{\text{Desired}}+ \underbrace{n_{i}(t)}_{\text{Noise}},
\end{align}
where $X_{I_2}(t)$ indicates transmitted signal from IoD to AA.

{{The instantaneous SNR at AA from IoD is 
\begin{align}
\gamma_6=\frac{P_{h3} \frac{\mid h_{i} \mid^2}{(d_{i})^\nu}}{ \sigma_i^2}.
\end{align}}}

\section{Outage Analysis and problem formulation}
%\subsection{Cellular network outage analysis}
%Cellular network outage can be computed as follows.
%\begin{multline}
%\mathscr{P}_{out}^{C} = \big(1 - \mathscr{P}[\Upsilon_{p}^1\geq \Upsilon_{c1}] \times \mathscr{P}[\Upsilon_{2} \geq \Upsilon_{c1}] \times \mathscr{P}[\Upsilon_{3}\geq \Upsilon_{c2}]\\ \times \mathscr{P}[\Upsilon_{4}\geq \Upsilon_{c2}]
%\big)
%\end{multline}
%where $\Upsilon_{c1}=2^{\frac{2R_c}{\tau}}-1$, $\Upsilon_{c2}=2^{\frac{2R_c}{1-\tau}}-1$, $R_c$ indicates the target rate of cellular communication.

{{\subsection{LU Network Outage Analysis}}}
The LU network outage can be expressed as 
{{\begin{multline}
\mathscr{P}_{out_C} = \big(1 - \mathscr{P}[\gamma_{p}\geq \gamma_{c1}] \times \mathscr{P}[\gamma_{2} \geq \gamma_{c2}] \times \mathscr{P}[\gamma_{3}\geq \gamma_{c}] \\ \times \mathscr{P}[\gamma_{4}\geq \gamma_{c}]
\big),
\end{multline}}}
{{where $\gamma_{c1}=2^{\frac{2R_{c1}}{\tau}}-1$, $\gamma_{c2}=2^{\frac{2R_{c2}}{1-\tau}}-1$, $\gamma_{c}=2^{{4R_c}}-1$}}. $R_{c1}$, ${R_{c2}}$ and ${R_{c}}$  indicate fixed transmission rates of LU communication in Phase-1, Phase-2 and Phase-3, respectively. {{Since, Phase-1 is related to the transmission from VLC system over VLC link and Phase-2 is related to transmission over RF link, therefore, the target rates are considered to be different for Phase-1 and Phase-2. In addition to that, the transmission in Phase-3 is related to relaying of information based on the available harvesting energy and harvesting energy is limited. As, the amount of harvesting energy is limited, therefore, it can support very small data rate compared to uplink and downlink transmission.}} 

{{$\mathscr{P}[\gamma_{p}\geq \gamma_{c1}]$}} is determined using the probability density function (PDF) of VLC link co-efficient. LED source is located at height
$L$ from AA and Euclidian distance $d_t$. $r_t$ is the radical distance as $\cos(\Theta)=\cos(\Phi)=\frac{L}{d_t}$ and $d_t=(L^2+r_t^2)^{1/2}$. Now, $h_{vc}$ can be found as [28]
\begin{equation}
h_{L}=\frac{\varepsilon (\varsigma+1) L^{\varsigma+1}}{(d_t)^{{\varsigma+3}}}
=\frac{\varepsilon (\varsigma+1) L^{\varsigma+1}}{(L^2+r_t^2)^{\frac{\varsigma+3}{2}}},
\end{equation}
where $\varepsilon=\frac{A_s}{2\pi}$. Since, we consider outdoor VLC link between VLC system and AA, therefore the link follows Nakagami fading with gamma distribution as follows : $\scalebox{01}{$f_{Z}{(z)}=\frac{n_m^{n_m} z^{n_m-1} \exp\big(-{\frac{z}{{1}/{n_m}}}\big)}{\Gamma(n_m)}$}$. Symbol $n_m$ is used to indicate the shaping factor of Nakagami fading model.

Now, {{$\mathscr{P}[\gamma_{p}\geq \gamma_{c1}]$}} can be expressed as [29]
 {{\begin{equation}
\mathscr{P}[\gamma_{p}\geq \gamma_{c1}]=\mathscr{P}\bigg[\gamma_{vc} \geq \frac{\gamma_{c1}}{\gamma_{s} (h_L)^2}\bigg]=\frac{\Gamma\big\{n_m ,n_m \frac{\gamma_{c1}}{\gamma_{s} (h_L)^2}\big\}}{\Gamma(n_m)},
\end{equation}}}
where {{$\gamma_{vc}=\mid h_{vc} \mid^2 $, $\gamma_s=\frac{(\vartheta \sqrt{N_L P_{L}} A_p)^2}{qI_nW}$}}, $\Gamma(.,.)$ indicates upper incomplete Gamma function.

$h_r$ and $h_i$ also follow Nakagami distribution. $h_{bt}$ follows Rayleigh distribution. $\mid h_r \mid^2=X_r$, $\mid h_{bt} \mid^2=X_{bt}$ and $\mid h_i \mid^2=X_i$. Now, 
{{\begin{equation}
\mathscr{P}[\gamma_{2} \geq \gamma_{c2}] =\frac{\Gamma\big(n_m ,n_m a_r\big)}{\Gamma(n_m)},
\end{equation}}}
%=\int_{a_r}^{\infty}  \frac{n_m^{n_m} z^{n_m-1} \exp\big(-{\frac{x_r}{{1}/{n_m}}}\big)}{\Gamma(n_m)} dx_r
where {{$a_r=\frac{\gamma_{c2} d_m^{\nu}\sigma_2^2}{(1-\rho)P_m}$}}.

Here, the VLC link is considered to be reciprocal in nature. Therefore, {{$\gamma_{rvc}$ can be written as $\gamma_{rvc}=\gamma_{vc}$. Now, $\mathscr{P}[\gamma_{3}\geq \gamma_{c}]$ can be written as  
\begin{multline} 
\mathscr{P}[\gamma_{3}\geq \gamma_{c}]=\mathscr{P}\bigg[\frac{{a_{1c}^\prime} (a_h \gamma_{vc} +b_h X_r) \gamma_{vc} }{{b_{1c}^\prime} (a_h \gamma_{vc} +b_h X_r) \gamma_{vc}  + 1} \geq \gamma_{c} \bigg] \\= \begin{cases}
\mathscr{P}\bigg[\gamma_{vc} \geq \frac{u_{1c}}{(a_h \gamma_{vc} +b_h X_r) } \bigg];\hspace*{0.3cm} \gamma_{c} < \frac{\varrho}{(1-\varrho)}\\
\scalebox{1}{$0;$} \hspace*{4cm} {{\text otherwise}}.\\ 
\end{cases}
\end{multline}}}

{{Now, $\mathscr{P}\bigg[\gamma_{vc} \geq \frac{u_{1c}}{(a_h \gamma_{vc} +b_h X_r) } \bigg]$ can be determined as [30]
\begin{multline}
\mathscr{P}\bigg[\gamma_{vc} \geq \frac{u_{1c}}{(a_h \gamma_{vc} +b_h X_r) } \bigg]=1-\frac{1}{\Gamma(n_m)}{\Upsilon\Bigg(n_m,\frac{\sqrt{\frac{u_{1c} }{a_{h}}}}{1/n_m}\Bigg)}\\+\frac{\sum_{p_a=0}^{n_m-1}({n_m})^{n_m+p_a+l}\sum_{r=0}^{p_a}(-1)^{r}{{p_a}\choose{r}} \big(\frac{u_{1c}}{b_{h}}\big)^{p_a-r} \big(\frac{a_{h}}{b_{h}}\big)^{r} }{p_a! l! (2r+ n_m -p_a-l+2t_l)\Gamma(n_m)}\\\sum_{l=0}^{\infty}(-1)^l \sum_{t_l=0}^{l}{{l}\choose{t_l}}     {\bigg(\frac{u_{1c} }{b_{h}}\bigg)}^{l-t_l} \bigg({1-\frac{a_{h}}{b_{h}}} \bigg)^{t_l}  \\\bigg(\sqrt{\frac{u_{1c}}{a_{h}}}\bigg)^{n_m+2r -p_a-l+2t_l},  
\end{multline}}}
where $a_h=\frac{2 f V_T(\vartheta \sqrt{N_L P_{L}}B_{dc})^2 {h_L}^2 \tau}{I_o }$, $b_h=\frac{2\eta \rho {P_{m}}{(1-\tau) }}{{(d_{m})^{\nu}}}$, {{$\Upsilon(.,.)$ indicates lower incomplete Gamma function}}.

{{$\mathscr{P}[\gamma_{4}\geq \gamma_{c}]$ can be written as
\begin{multline} 
\mathscr{P}[\gamma_{4}\geq \gamma_{c}]=\mathscr{P}\bigg[\frac{{a_{2c}^\prime} (a_h \gamma_{vc} +b_h X_r) X_r }{{b_{2c}^\prime} (a_h \gamma_{vc} +b_h X_r) X_r  + 1} \geq \gamma_{c} \bigg] \\= \begin{cases}
\mathscr{P}\bigg[X_r \geq \frac{u_{2c}}{(a_h \gamma_{vc} +b_h X_r) } \bigg];\hspace*{0.3cm} \gamma_{c} < \frac{\varrho}{(1-\varrho)}\\
\scalebox{1}{$0;$} \hspace*{4cm} {{\text otherwise}}.\\ 
\end{cases}
\end{multline}

Now, $\mathscr{P}\bigg[X_r\geq \frac{u_{2c}}{(a_h \gamma_{vc} +b_h X_r) } \bigg]$ can be determined as follows.
\begin{multline}
\mathscr{P}\bigg[X_r \geq \frac{u_{2c}}{(a_h \gamma_{vc} +b_h X_r) } \bigg]=1-\frac{1}{\Gamma(n_m)}{\Upsilon\Bigg(n_m,\frac{\sqrt{\frac{u_{2c} }{b_{h}}}}{1/n_m}\Bigg)}\\+\frac{\sum_{p_a=0}^{n_m-1}({n_m})^{n_m+p_a+l}\sum_{r=0}^{p_a}(-1)^{r}{{p_a}\choose{r}} \big(\frac{u_{2c}}{a_{h}}\big)^{p_a-r} \big(\frac{b_{h}}{a_{h}}\big)^{r} }{p_a! l! (2r+ n_m -p_a-l+2t_l)\Gamma(n_m)}\\\sum_{l=0}^{\infty}(-1)^l \sum_{t_l=0}^{l}{{l}\choose{t_l}}     {\bigg(\frac{u_{2c} }{a_{h}}\bigg)}^{l-t_l} \bigg({1-\frac{b_{h}}{a_{h}}} \bigg)^{t_l}  \\\bigg(\sqrt{\frac{u_{2c}}{b_{h}}}\bigg)^{n_m+2r -p_a-l+2t_l}, 
\end{multline}
where $u_{1c}=\frac{\gamma_{c}}{a_{1c}^\prime-\gamma_{c} b_{1c}^\prime}$, $u_{2c}=\frac{\gamma_{c}}{a_{2c}^\prime-\gamma_{c} b_{2c}^\prime}$}}, $a_{1c}^\prime=\frac{{\vartheta^2 h_L^2 \varrho   A_p^2}}{\sigma_1^2}$, $b_{1c}^\prime=\frac{\vartheta^2 h_L^2 (1-\varrho)   A_p^2}{\sigma_1^2}$, $a_{2c}^\prime=\frac{\varrho}{(d_{m})^\nu \sigma_2^2}$, $b_{2c}^\prime=\frac{1-\varrho}{(d_{m})^\nu \sigma_2^2}$.

Finally, the closed form outage expression of LU communication can be found as (26) (as shown at the top of the next page) using (20)-(21), (23), (25).
\begin{figure*}
{{\begin{multline}
\mathscr{P}_{out_C}=1-\Bigg[\Bigg\{\frac{\Gamma\big\{n_m ,n_m \frac{\gamma_{c1}}{\gamma_{s} (h_L)^2}\big\}}{\Gamma(n_m)}\Bigg\}\times \Bigg\{ \frac{\Gamma\big(n_m ,n_m a_r\big)}{\Gamma(n_m)}\Bigg\}\times\Bigg\{1-\frac{1}{\Gamma(n_m)}{\Upsilon\Bigg(n_m,\frac{\sqrt{\frac{u_{1c} }{a_{h}}}}{1/n_m}\Bigg)}+\frac{\sum_{p_a=0}^{n_m-1}({n_m})^{n_m+p_a+l}\sum_{r=0}^{p_a}(-1)^{r}{{p_a}\choose{r}}  }{p_a! l! (2r+ n_m -p_a-l+2t_l)}\\\frac{\big(\frac{u_{1c}}{b_{h}}\big)^{p_a-r} }{\Gamma(n_m)}\Big(\frac{a_{h}}{b_{h}}\Big)^{r}\sum_{l=0}^{\infty}(-1)^l \sum_{t_l=0}^{l}{{l}\choose{t_l}}     {\bigg(\frac{u_{1c} }{b_{h}}\bigg)}^{l-t_l} \bigg({1-\frac{a_{h}}{b_{h}}} \bigg)^{t_l}  \bigg(\sqrt{\frac{u_{1c}}{a_{h}}}\bigg)^{n_m+2r -p_a-l+2t_l}  \Bigg\}\times\Bigg\{1-\frac{1}{\Gamma(n_m)}{\Upsilon\Bigg(n_m,\frac{\sqrt{\frac{u_{2c} }{b_{h}}}}{1/n_m}\Bigg)}\\+\frac{\sum_{p_a=0}^{n_m-1}({n_m})^{n_m+p_a+l}\sum_{r=0}^{p_a}(-1)^{r}{{p_a}\choose{r}} \big(\frac{u_{2c}}{a_{h}}\big)^{p_a-r} \big(\frac{b_{h}}{a_{h}}\big)^{r} }{p_a! l! (2r+ n_m -p_a-l+2t_l)\Gamma(n_m)}\sum_{l=0}^{\infty}(-1)^l \sum_{t_l=0}^{l}{{l}\choose{t_l}}     {\bigg(\frac{u_{2c} }{a_{h}}\bigg)}^{l-t_l} \bigg({1-\frac{b_{h}}{a_{h}}} \bigg)^{t_l}  \bigg(\sqrt{\frac{u_{2c}}{b_{h}}}\bigg)^{n_m+2r -p_a-l+2t_l}   \Bigg\}\Bigg].
\end{multline}}}
\hrulefill
\end{figure*}

\subsection{IoT Network Outage Analysis}
The IoT network outage can be determined as follows.
{{\begin{multline}
\mathscr{P}_{out_I} = \big(1 - \mathscr{P}[\gamma_{p}\geq \gamma_{c1}] \times \mathscr{P}[\gamma_{2} \geq \gamma_{c2}] \times \mathscr{P}[\gamma_{5}\geq \gamma_{i}] \\ \times \mathscr{P}[\gamma_{6}\geq \gamma_{i}]
\big),
\end{multline}
where $\gamma_{i}=2^{{4R_i}}-1$}}, $R_i$ indicates the fixed transmission rate of IoT communication.

{{$\mathscr{P}[\gamma_{5}\geq \gamma_{i}]$ can be determined as
\begin{equation}
\mathscr{P}[\gamma_{5}\geq \gamma_{i}]=\mathscr{P}[(a_h \gamma_{vc} +b_h X_r)c_s X_i\geq \gamma_{i}].
\end{equation}}}

Now, (28) can be determined as (29) [29, Sec. 3.381.2], [30] with $c_s=\frac{(1-\varrho)   }{ \sigma_i^2 (d_{i})^\nu}$.
\begin{figure*}
{{\begin{multline}
\mathscr{P}[(a_h \gamma_{vc} +b_h X_r)c_s X_i\geq \gamma_{i}]
= \sum_{p_a=0}^{n_m-1}  \frac{\big(\frac{n_m}{b_h}\big)^{p_a}\big(\frac{\gamma_{i}}{c_s}\big)^{p_a-r} ({a_{h}})^{r}}{p_a!  ({1/n_m})^{ n_m} \Gamma(n_m)} \sum_{r=0}^{p_a}(-1)^{r}{{p_a}\choose{r}}\frac{\Gamma{(n_m +r)}}{({1/n_m})^{n_m} \Gamma(n_m){\big(n_m-\frac{a_{h}n_m}{b_{h}}\big)}^{n_m+r}}\\ \bigg[2\bigg(\frac{\gamma_{i} }{c_s b_{h} } \bigg)^{\frac{(n_m-p_a+r)}{2}} K_{n_m-p_a+r}\bigg(2n_m\sqrt{\frac{\gamma_{i} }{c_s b_{h}}}\bigg)   - \frac{\sum_{j=0}^{n_m+r-1}  \bigg[\frac{\gamma_{i}  \big({n_m}-\frac{a_{h}{n_m}}{b_{h}}\big)}{c_s a_{h}} \bigg]^{j}  2\big[\frac{\gamma_{i} }{c_s a_{h} }\big]^{\frac{(n_m-p_a+r-j)}{2}}}{j!} K_{n_m-p_a+r-j}\bigg(2  {n_m}\sqrt{\frac{\gamma_{i} }{c_s a_{h}}}\bigg)  \bigg]\\+\frac{\sum_{p_a=0}^{n_m-1}2\big({\frac{\gamma_{i} {n_m}}{c_s a_{h}}}\big)^{p_a}  ({n_m})^{n_m}{\big(      {\frac{\gamma_{i} }{c_s a_{h} }}\big)}^{\frac{(n_m - p_a)}{2}}K_{n_m - p_a}\Big(2n_m\sqrt{{{\frac{\gamma_{i} }{c_s a_{h} }}}}\Big)}{p_a! \Gamma(n_m)}.
\end{multline} }}
\hrulefill
\end{figure*}
%=\mathscr{P}\bigg[\Upsilon_{vc} \geq \frac{\Upsilon_{i}}{c_s a_h X_i}-\frac{b_h X_r}{a_h}\bigg]

{{$\mathscr{P}[\gamma_{6}\geq \gamma_{i}]$ can be determined as [29, Sec. 3.324.1]
\begin{multline}
\hspace*{-0.08cm} \mathscr{P}[\gamma_{6}\geq \gamma_{i}]=\int_{0}^{\infty}   \frac{n_m^{n_m} x_i^{n_m-1} }{\Gamma(n_m)}  \exp\bigg(-{\frac{x_i}{{1}/{n_m}}}-\frac{a_i}{x_i}\bigg)  dx_i \\= 2\frac{n_m^{n_m} }{\Gamma(n_m)} \bigg(\frac{a_i}{n_m}\bigg)^{\frac{n_m}{2}}\mathscr{K}_{n_m}\big(2\sqrt{a_i n_m}\big),
\end{multline}
where $a_i=\frac{\gamma_{i} }{a_{bt}}$}}, $a_{bt}=\frac{\eta   P_{bt}}{(d_{bt})^{\nu}}  \frac{3}{ d_i^{\nu} \sigma_i^2}$ and $\mathscr{K}_{\varpi}(.)$ indicates $\varpi^{th}$ order modified Bessel function.

Finally, the closed form outage expression of IoT communication can be found as (31) (as shown at the top of the next page) using (20)-(21), (29)-(30). 
\begin{figure*}
{{\begin{multline}
\mathscr{P}_{out_I} =1-\Bigg[\Bigg\{\frac{\Gamma\big\{n_m ,n_m \frac{\gamma_{c1}}{\gamma_{s} (h_L)^2}\big\}}{\Gamma(n_m)}\Bigg\}\times \Bigg\{ \frac{\Gamma\big(n_m ,n_m a_r\big)}{\Gamma(n_m)}\Bigg\}\times \Bigg\{\sum_{p_a=0}^{n_m-1}  \frac{\big(\frac{n_m}{b_h}\big)^{p_a}\big(\frac{\gamma_{i}}{c_s}\big)^{p_a-r} ({a_{h}})^{r}}{p_a!  ({1/n_m})^{ n_m} \Gamma(n_m)}\frac{ \sum_{r=0}^{p_a}(-1)^{r}{{p_a}\choose{r}}\Gamma{(n_m +r)}}{({1/n_m})^{n_m} \Gamma(n_m){\big(n_m-\frac{a_{h}n_m}{b_{h}}\big)}^{n_m+r}}\\ \bigg[2\bigg(\frac{\gamma_{i} }{c b_{h} } \bigg)^{\frac{(n_m-p_a+r)}{2}} K_{n_m-p_a+r}\bigg(2n_m\sqrt{\frac{\gamma_{i} }{c b_{h}}}\bigg)   - \frac{\sum_{j=0}^{n_m+r-1}  \bigg[\frac{\gamma_{i}  \big({n_m}-\frac{a_{h}{n_m}}{b_{h}}\big)}{c_s a_{h}} \bigg]^{j}  2\big[\frac{\gamma_{i} }{c_s a_{h} }\big]^{\frac{(n_m-p_a+r-j)}{2}}}{j!} K_{n_m-p_a+r-j}\bigg(2{n_m}\sqrt{\frac{\gamma_{i}  }{c_s a_{h}}}\bigg)  \bigg]\\+\frac{\sum_{p_a=0}^{n_m-1}2\big({\frac{\gamma_{i} {n_m}}{c_s a_{h}}}\big)^{p_a}  ({n_m})^{n_m}{\big(      {\frac{\gamma_{i} }{c_s a_{h} }}\big)}^{\frac{(n_m - p_a)}{2}}K_{n_m - p_a}\Big(2n_m\sqrt{{{\frac{\gamma_{i} }{c_s a_{h} }}}}\Big)}{p_a! \Gamma(n_m)}\Bigg\} \times \Bigg\{ 2\frac{n_m^{n_m} }{\Gamma(n_m)} \bigg(\frac{a_i}{n_m}\bigg)^{\frac{n_m}{2}}\mathscr{K}_{n_m}\big(2\sqrt{a_i n_m}\big)\Bigg\} \Bigg].
\end{multline}}}
\hrulefill
\end{figure*}

%where $C_{\varepsilon}=\frac{\big[\varepsilon (\varsigma+1) L^{\varsigma+1}\big]^{\frac{2}{\varsigma+3}} \sum_{t_k=0}^\infty (-1)^{t_k} \sum_{l=0}^{t_k}  {t_k \choose l} \sum_{t_l=0}^l  {l \choose t_l}}{r_m^2 (\varsigma+3)(l-2t_l+1)t_k!\big(t_k+2t_l-2l-\frac{\varsigma+4}{\varsigma+3}\big)} $    
%$\{\varepsilon (\varsigma+1) L^{\varsigma+1}\}^{2\big(t_k+2t_l-2l-\frac{\varsigma+4}{\varsigma+3}\big)} \bigg[{\frac{1}{L^{2}}}-\frac{1}{r_m^2+L^2}\bigg]^{\frac{t_k+2t_l-2l-\frac{\varsigma+4}{\varsigma+3}}{\varsigma+3}}$.

\subsection{Throughput Analysis}
The throughput of LU network and IoT network can be expressed as  
\begin{equation}
T_C=2 \times (1-\mathscr{P}_{out_C})\times R_c \times \frac{T}{4T},
\end{equation}

\begin{equation}
T_I=2 \times (1-\mathscr{P}_{out_I}) \times R_i \times \frac{T}{4T}.
\end{equation}

\subsection{High SNR Approximation}
{{At high SNR region, closed form expressions of LU outage and IoT outage can be approximated to (34) and (35), respectively.}} In case of high SNR approximation, $\Gamma(N,x)\approx \Gamma(N)\exp(-x)$ (for, $x$ $<<$ 1), $\exp(-x) \approx 1-x$ and $\mathscr{K}_{\varpi}(x) \approx \frac{1}{2}\Gamma(\varpi){\big(\frac{2}{x}\big)}^{\varpi}$, $\varpi >$ 0 [29].

%At high SNR region, $t_k$ is going from 0 to 1.
 Now, the LU outage expression is approximated as 
\begin{multline}
\hspace{-0.2cm}\mathscr{P}_{out_C}^\infty  \simeq 1-\Bigg[\Big\{1- \frac{n_m\gamma_{c1}}{\gamma_{s} (h_L)^2}\Big\}\times (1-n_m a_r)\times \Big\{n_m \sqrt{\frac{u_{1c} }{a_{h}}}+\\ C_p {\bigg({\frac{1}{ b_{h}}\bigg)} }^{{p_a+l}} {\bigg({\frac{1}{a_h }\bigg)} }^{\frac{n_m-p_a-l+2t_l}{2}}   \bigg({b_h-{a_{h}}} \bigg)^{t_l}(\sqrt{{u_{1c}}})^{n_m+p_a+l}\Big\}\\ \times \Big\{n_m \sqrt{\frac{u_{2c} }{b_{h}}} +C_p {\bigg({\frac{1}{ a_{h}}\bigg)} }^{{p_a+l}} {\bigg({\frac{1}{b_h }\bigg)} }^{\frac{n_m-p_a-l+2t_l}{2}}   \bigg({a_h-{b_{h}}} \bigg)^{t_l}\\(\sqrt{{u_{2c}}})^{n_m+p_a+l}\Big\}\Bigg],
\end{multline}
where $\scalebox{0.87}{$C_p=\frac{\sum_{p_a=0}^{n_m-1}({n_m})^{n_m+p_a+l}\sum_{r=0}^{p_a}(-1)^{r}{{p_a}\choose{r}} \sum_{l=0}^{1}(-1)^{l}\sum_{t_l=0}^{l}{{l}\choose{t_l}}}{p_a! l!  (2r+ n_m -p_a-l+2t_l)\Gamma(n_m)}
  $}$.

At high SNR, the IoT outage expression is approximated as 
\begin{multline}
\hspace{-0.2cm}\mathscr{P}_{out_I}^\infty  \simeq  1- \Bigg\{1-\frac{n_m \gamma_{c1}}{\gamma_{s} (h_L)^2}\Bigg\} \times (1-n_m a_r)   \times \Bigg\{\frac{C_i({a_{h}b_h})^{r}}{{\big(b_h-{a_{h}}\big)}^{n_m+r} } \\(b_h)^{n_m-p_a}\Big[\Gamma(n_m-p_a+r)-\frac{\Gamma(n_m-p_a+r-j)\sum_{j=0}^{n_m+r-1} }{j!}\\\Big({\frac{\gamma_{i} n_m^{2}}{c_s a_{h} b_h }}\Big)^j\Big(b_h-{a_h}\Big)^j\Big]+\frac{\sum_{p_a=0}^{n_m-1}\big({\frac{\gamma_{i} {n_m}}{c_s a_{h}}}\big)^{p_a}  ({n_m})^{p_a}}{p_a! \Gamma(n_m)}\Bigg\},
\end{multline}
where $C_i=\sum_{p_a=0}^{n_m-1} \bigg\{ \frac{(n_m)^{2(p_a-r)}\big(\frac{\gamma_{i}}{c_s}\big)^{p_a-r}  \sum_{r=0}^{p_a}(-1)^{r}{{p_a}\choose{r}}}{p_a!   \{\Gamma(n_m)\}^2}\bigg\}$ 

\vspace*{0.25 cm}
$\hspace*{5.2 cm} \times \Gamma{(n_m +r)}$.
%where $C_{ab}=\Big\{\frac{f V_T(\vartheta \sqrt{N_L P_{L}}B_{dc})^2 }{I_o}\Big\}^{t_k-2l+2t_l-1}{\Big\{\frac{\eta \rho {P_{m}}}{{(d_{m})^{\nu}} }\Big\}}^{l-t_k-t_l}$.

\subsection{Problem Formulation}
\textbf{Problem-1}: Minimization of IoT network outage ($\mathscr{P}_{out_I}$), under the constraints of LU network outage threshold ($\mathscr{P}_{c_{th}}$), threshold of power allocation factor ($\varrho_{th}$), maximum transmission power limit of PSe, LED array of VT and power transmitter ($P_{tm}$, $P_{tL}$ and $P_{tk}$) and target rate threshold of LU communication ($R_{tc}$).

\begin{equation}
\begin{aligned}
& \underset{\tau}{\text{min}}
& & {\mathscr{P}_{out_I}} \\
& \text{s.t.} & & C_1 : \mathscr{P}_{out_C} \leq \mathscr{P}_{c_{th}},\\
& & &  C_2 :  \varrho \geq \varrho_{th},\\
& & &  C_3 : P_m \leq P_{tm}, P_L \leq P_{tL}, P_{bt} \leq P_{tk},  \\
& & &  C_4 : R_{c} \geq R_{tc}.
\end{aligned}
\end{equation}

\textbf{Problem-2}: Minimization of LU network outage ($\mathscr{P}_{out_C}$), under 
the constraints of IoT network outage threshold ($\mathscr{P}_{i_{th}}$), threshold of power allocation factor ($\varrho_{th}$), maximum transmission power limit of PSe and LED array of VT ($P_{tm}$ and $P_{tL}$), target rate threshold of IoT communication ($R_{ti}$).

\begin{equation}
\begin{aligned}
& \underset{\tau}{\text{min}}
& & {\mathscr{P}_{out_C}} \\
& \text{s.t.} & & C_1 : \mathscr{P}_{out_I} \leq \mathscr{P}_{i_{th}}, \\
& & &  C_2 :  \varrho \geq \varrho_{th},\\
& & &  C_3 : P_m \leq P_{tm}, P_L \leq P_{tL}, \\
& & &  C_4 : R_i \geq R_{ti}.
\end{aligned}
\end{equation}
%& \tau \geq \frac{T}{3}    \\

 Due to the non-linear nature of the first derivative of both Problem-1 and Problem-2, it is not possible to determine the closed form optimal solution of $\tau$. Therefore, the range of $\tau$ is possible to obtain through the one-way exhaustive searching technique.

%$r_m$=5 m, 
\section{Numerical results}
The necessary parameters and their considered values are: $\Theta_{1/2}=20^{\circ}$, $A_s=0.04$ m$^2$, {{$L=3$ m, $r_t=(0,2)$ m, $d_t=(3,\sqrt{13}$) m [6]}}, $B_{dc}=10$ mA, $\vartheta=0.4$ A/W, 
$\nu=2.7$, {{$\sigma_1^2=10^{-21}$ W}}, $A_p=10$ mA, $f=0.75$, $I_o=10^{-9}$ A, $P_m=1$ W (Fig. 3 to Fig. 6),
{{$\sigma_2^2=8$ $\mu$W}}, $\rho=0.9$, $\tau=0.1$, $\eta=0.9$, $P_{bt}=1$ W (Fig. 4 to Fig. 8), $d_m=4$ m, $\varrho=0.85$, $d_i=6$ m, $d_{bt}=10$ m, $\sigma_i^2=3$ $\mu$W. Monte-Carlo simulation is executed to verify the validity of analytical output. The LU outage threshold and and IoT outage threshold are considered as $\mathscr{P}_{c_{th}}=0.1$ and $\mathscr{P}_{i_{th}}=0.2$, respectively.  
 %\textdegree
%, $R_{c1}$=0.5 bps/Hz, $R_{c2}$=0.4 bps/Hz, $R_{c}$=0.4 bps/Hz, $R_{ti}$=0.25 bps/Hz 
  \begin{figure}[h]
  \centering
  \includegraphics[width=0.86\linewidth]{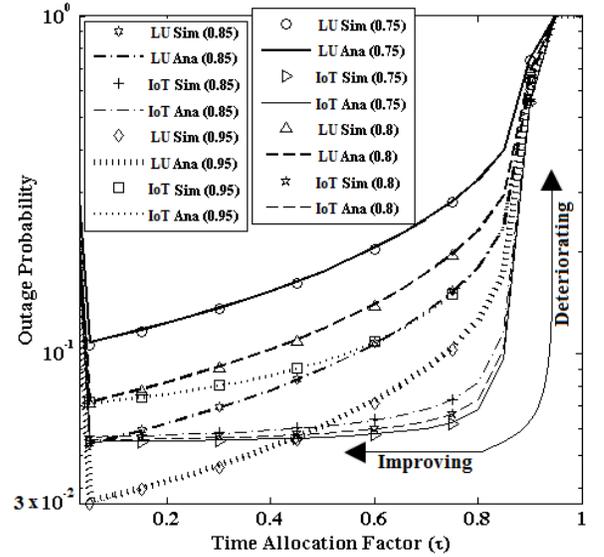}
  \caption{{{Outage probability vs $\tau$ for different values of $\varrho$ (values of $\varrho$ are given within bracket)}}}
 \end{figure}
  %\begin{figure}[h]
 % \centering
%  \includegraphics[width=0.83\linewidth]{plot_3}
  %\caption{\textbf{Outage probability vs $\tau$ for different values of $B_{dc}$, $N_l$ and $R_c$ }}
% \end{figure}
\begin{figure}[h]
  \centering
  \includegraphics[width=0.85\linewidth]{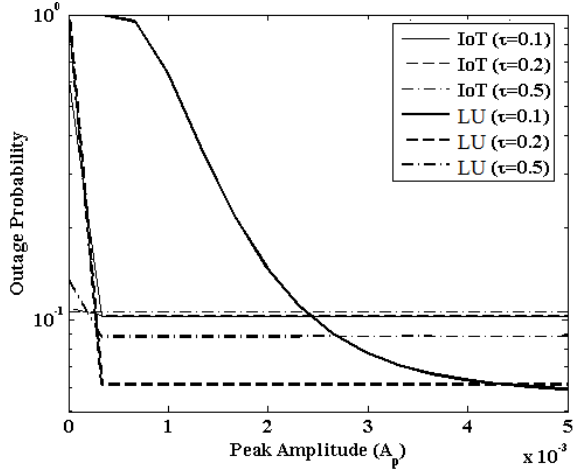}
  \caption{{Outage probability vs $A_p$}}
 \end{figure}

Fig. 3 illustrates the outage performances of both LU and IoT network with respect to the variation of $\tau$ for different values of $\varrho$ {{(considering $\varrho_{th}=0.7$)}} with $N_L=15$, $P_{L}=150$ mW, $P_{bt}=2$ W, $R_{c1}={R_{c2}}={R_{c}}=0.4$ bps and $R_i=0.25$ bps. Simulation results match analytical outputs properly. Due to more power allocation for relaying LU information and less power allocation for IoT communication, LU outage performance improves and IoT outage performance deteriorates with increasing $\varrho$. As shown in this figure, it is also observed that initially, the outage performances of both LU and IoT networks are improved with increasing $\tau$ and after reaching the minimum value, the outage performances are degraded with the further increase in $\tau$.
 
\begin{figure*}[t]
     %\centering
     \begin{subfigure}[b]{0.325\textwidth}
         \centering
         \includegraphics[width=\textwidth]{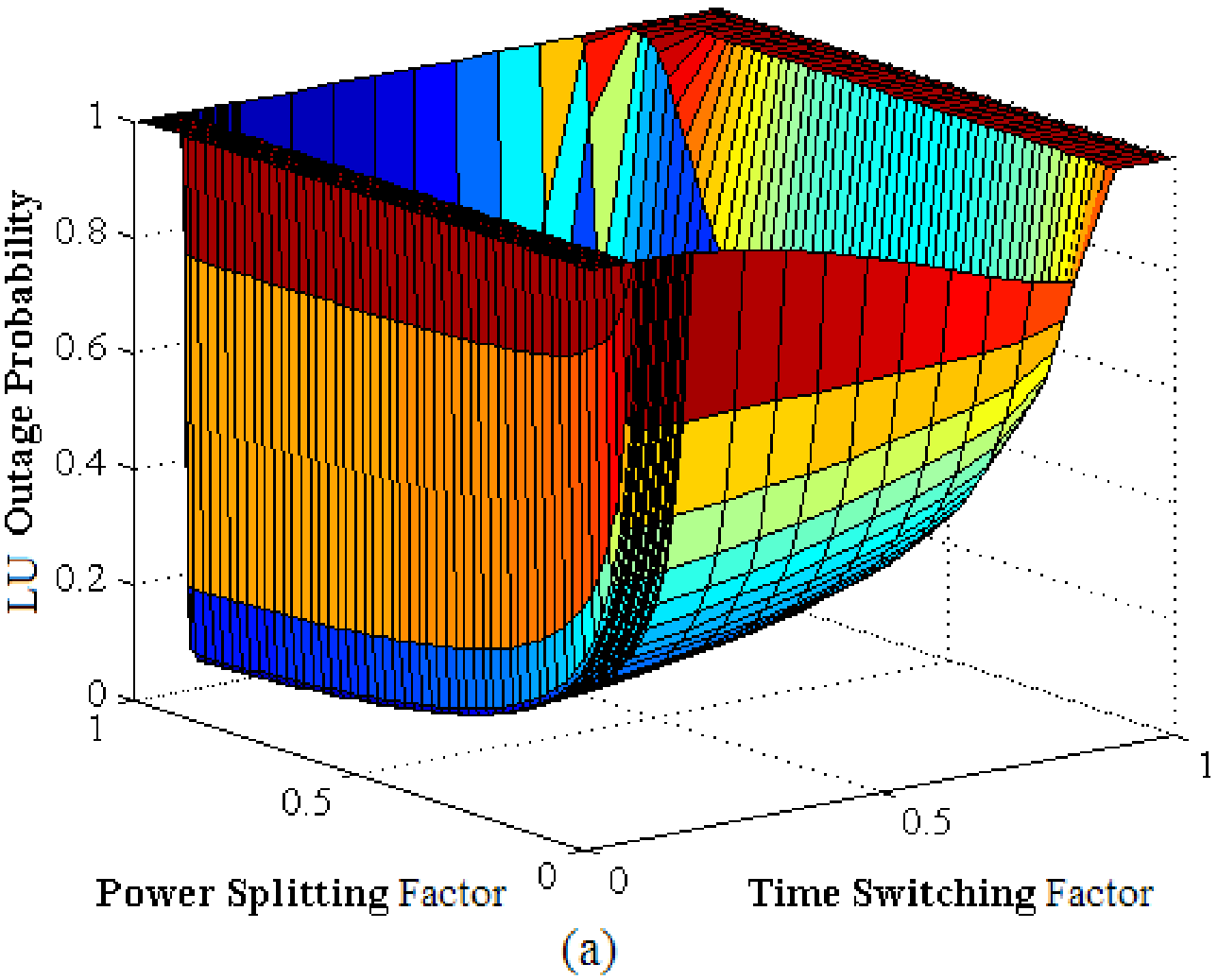}
        % \caption{\textbf{Outage probability vs $\lambda_1$=$\lambda_2$=$\lambda$ for different values of N$_{tb}$ and m$_k$=1 using hybrid relaying,}}
          \end{subfigure}
     \hfill
     \begin{subfigure}[b]{0.325\textwidth}
         \centering
         \includegraphics[width=\textwidth]{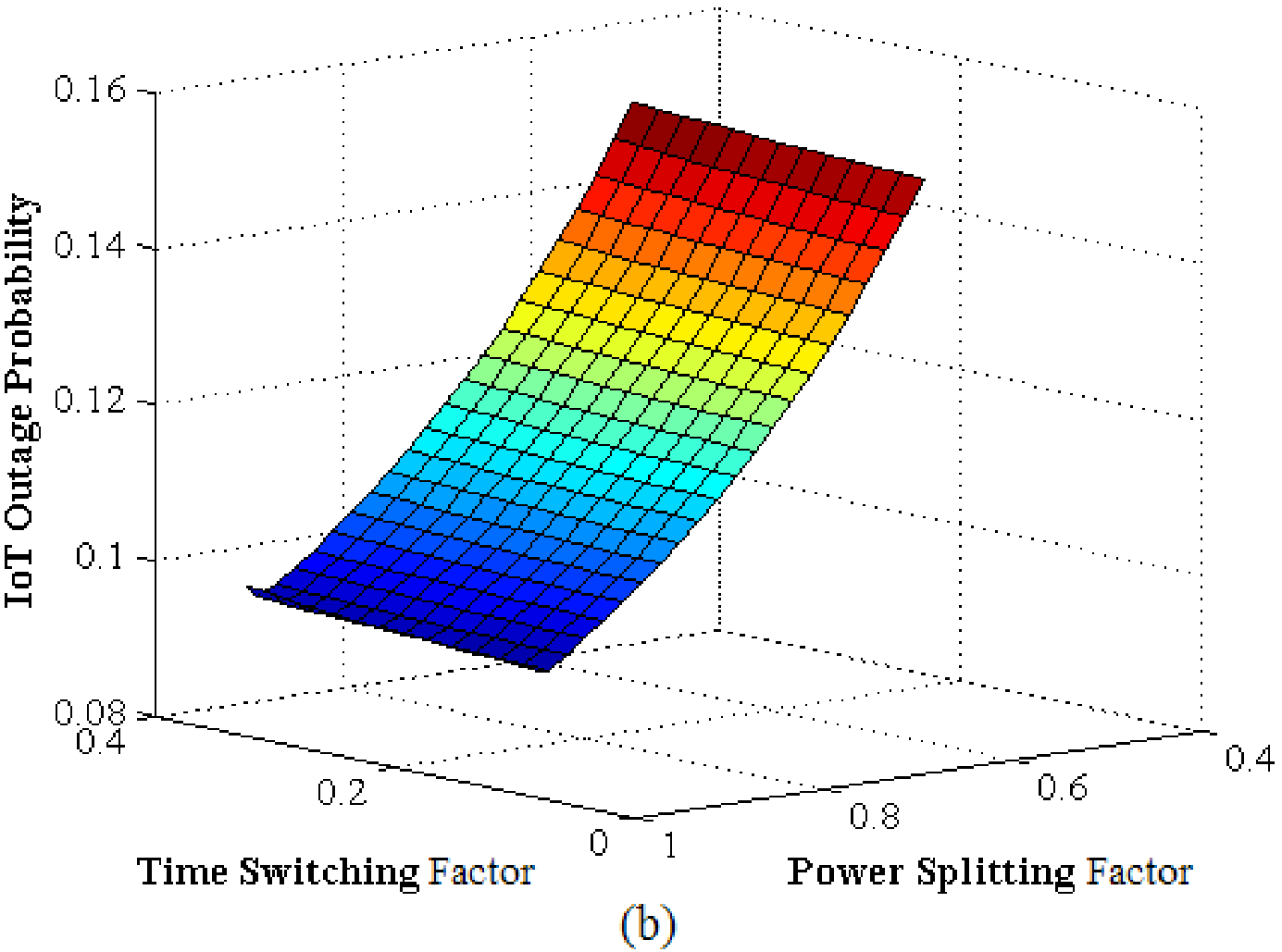}
         %\caption{\textbf{for different relaying techniques using N$_{tb}$=m$_k$=2}}
              \end{subfigure}
                \hfill
     \begin{subfigure}[b]{0.325\textwidth}
         \centering
         \includegraphics[width=\textwidth]{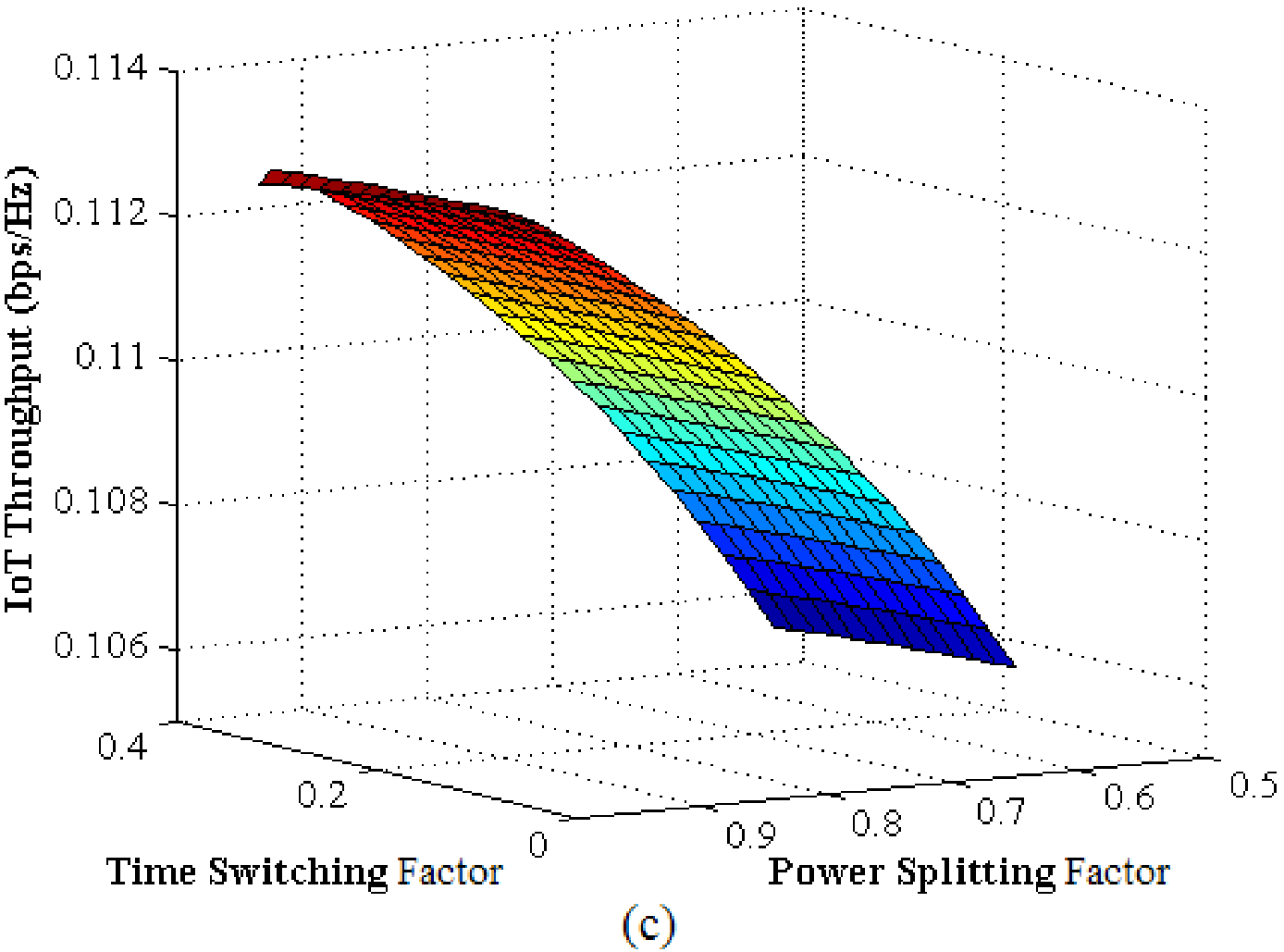}
         %\caption{\textbf{for different relaying techniques using N$_{tb}$=m$_k$=2}}
              \end{subfigure}
              \caption{{Outage probability vs $\rho$ vs $\tau$, (a) LU network, (b) IoT network, (c) IoT Throughput vs $\rho$ vs $\tau$}}
     \hfill
    \end{figure*}
    \begin{figure*}
     %\centering
     \begin{subfigure}[b]{0.325\textwidth}
         \centering
         \includegraphics[width=\textwidth]{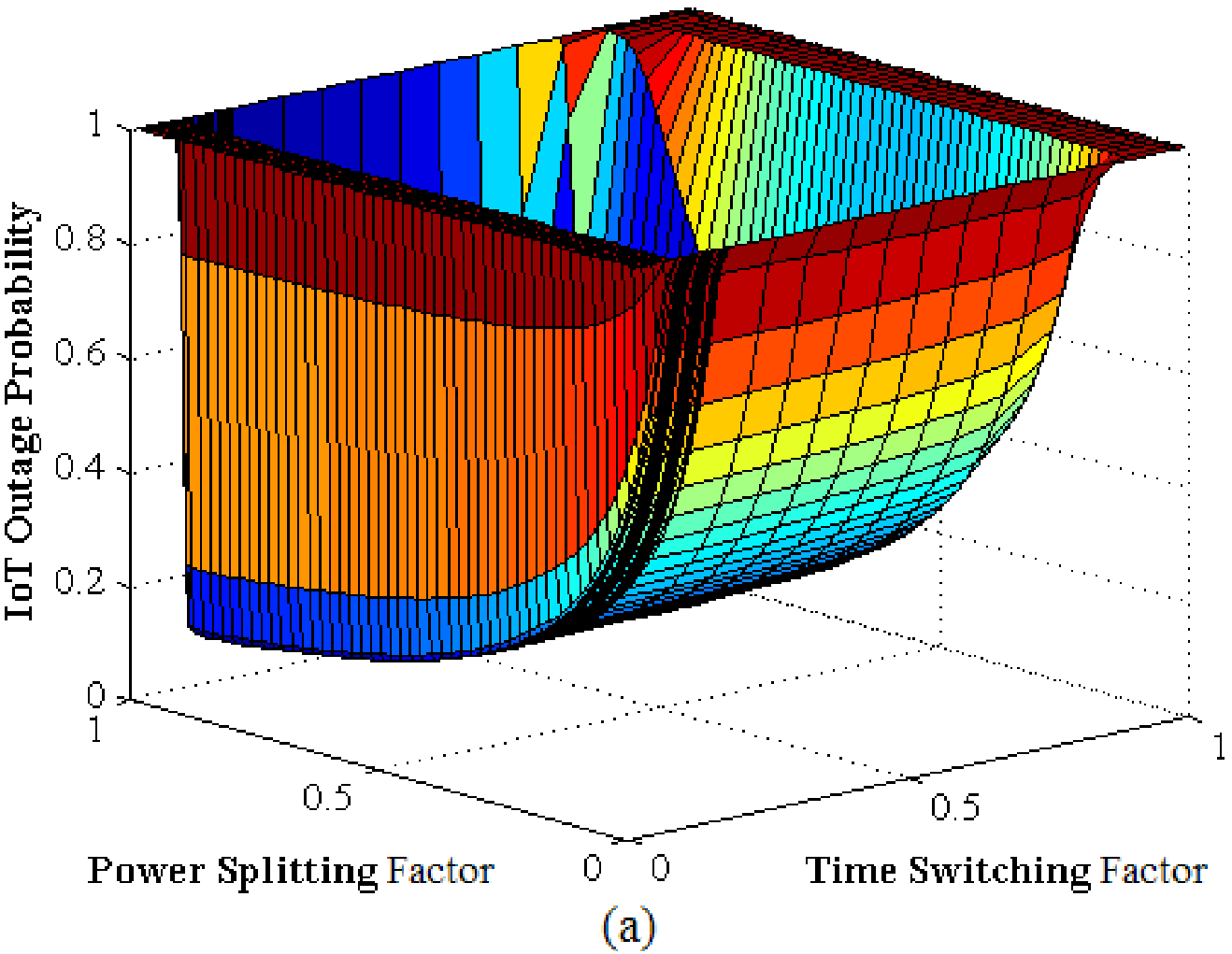}
        % \caption{\textbf{Outage probability vs $\lambda_1$=$\lambda_2$=$\lambda$ for different values of N$_{tb}$ and m$_k$=1 using hybrid relaying,}}
          \end{subfigure}
     \hfill
     \begin{subfigure}[b]{0.325\textwidth}
         \centering
         \includegraphics[width=\textwidth]{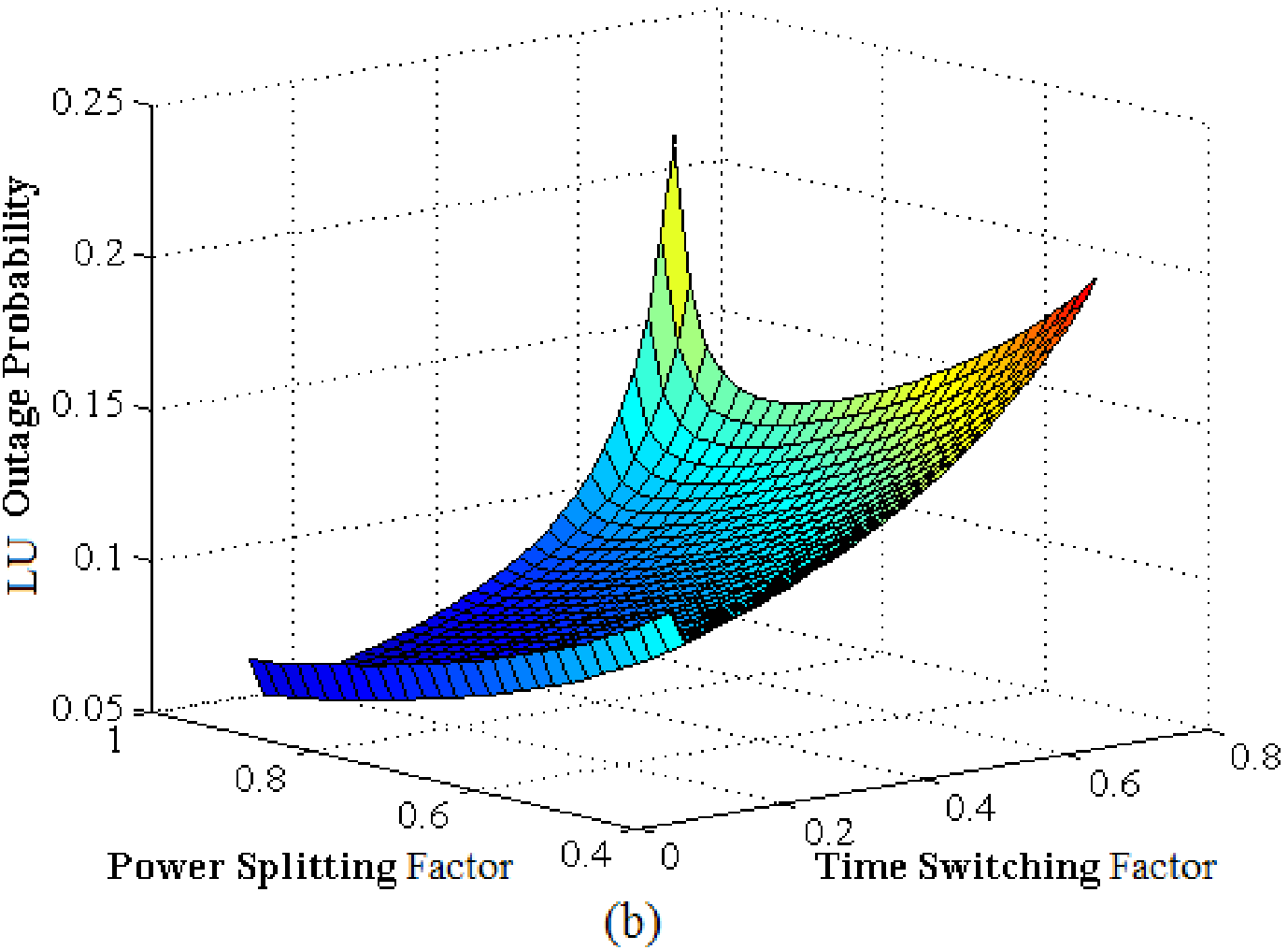}
         %\caption{\textbf{for different relaying techniques using N$_{tb}$=m$_k$=2}}
              \end{subfigure}
                \hfill
     \begin{subfigure}[b]{0.325\textwidth}
         \centering
         \includegraphics[width=\textwidth]{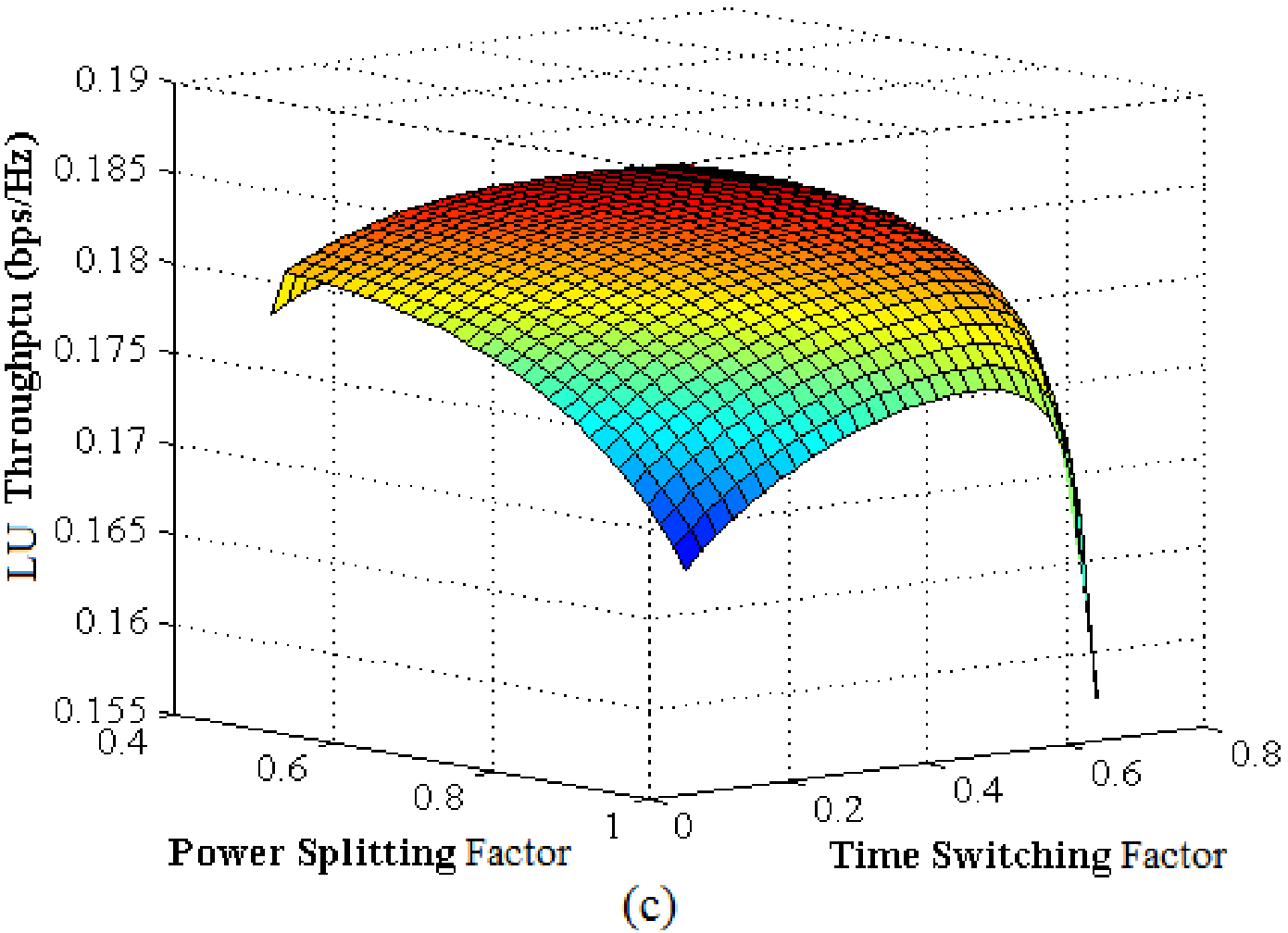}
         %\caption{\textbf{for different relaying techniques using N$_{tb}$=m$_k$=2}}
              \end{subfigure}
              \caption{{Outage probability vs $\rho$ vs $\tau$, (a) IoT network, (b) LU network, (c) LU Throughput vs $\rho$ vs $\tau$}}
     \hfill
    \end{figure*}
    
   \begin{figure*}[h]
     %\centering
     \begin{subfigure}[b]{0.325\textwidth}
         \centering
         \includegraphics[width=\textwidth]{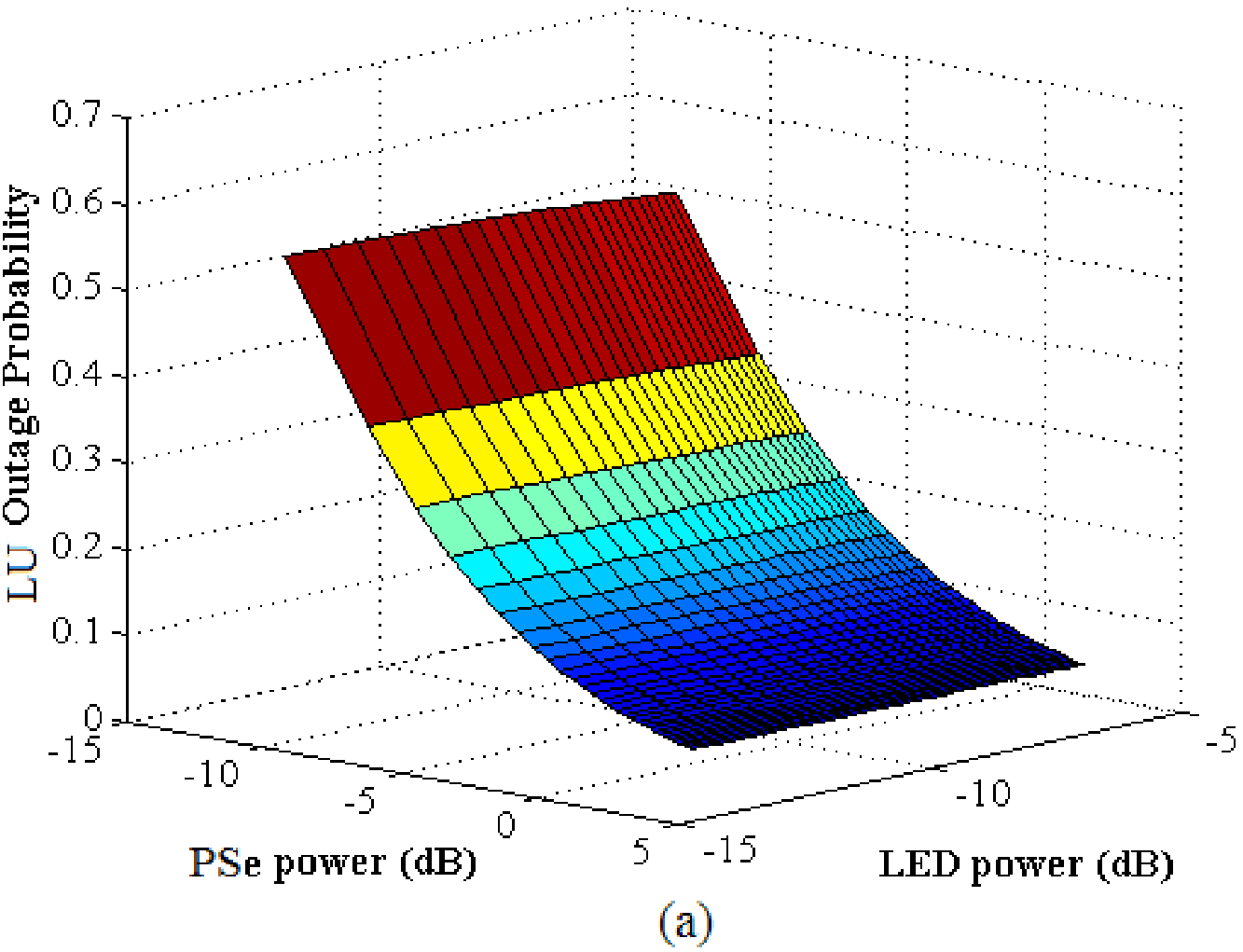}
        % \caption{\textbf{Outage probability vs $\lambda_1$=$\lambda_2$=$\lambda$ for different values of N$_{tb}$ and m$_k$=1 using hybrid relaying,}}
          \end{subfigure}
     \hfill
     \begin{subfigure}[b]{0.33\textwidth}
         \centering
         \includegraphics[width=\textwidth]{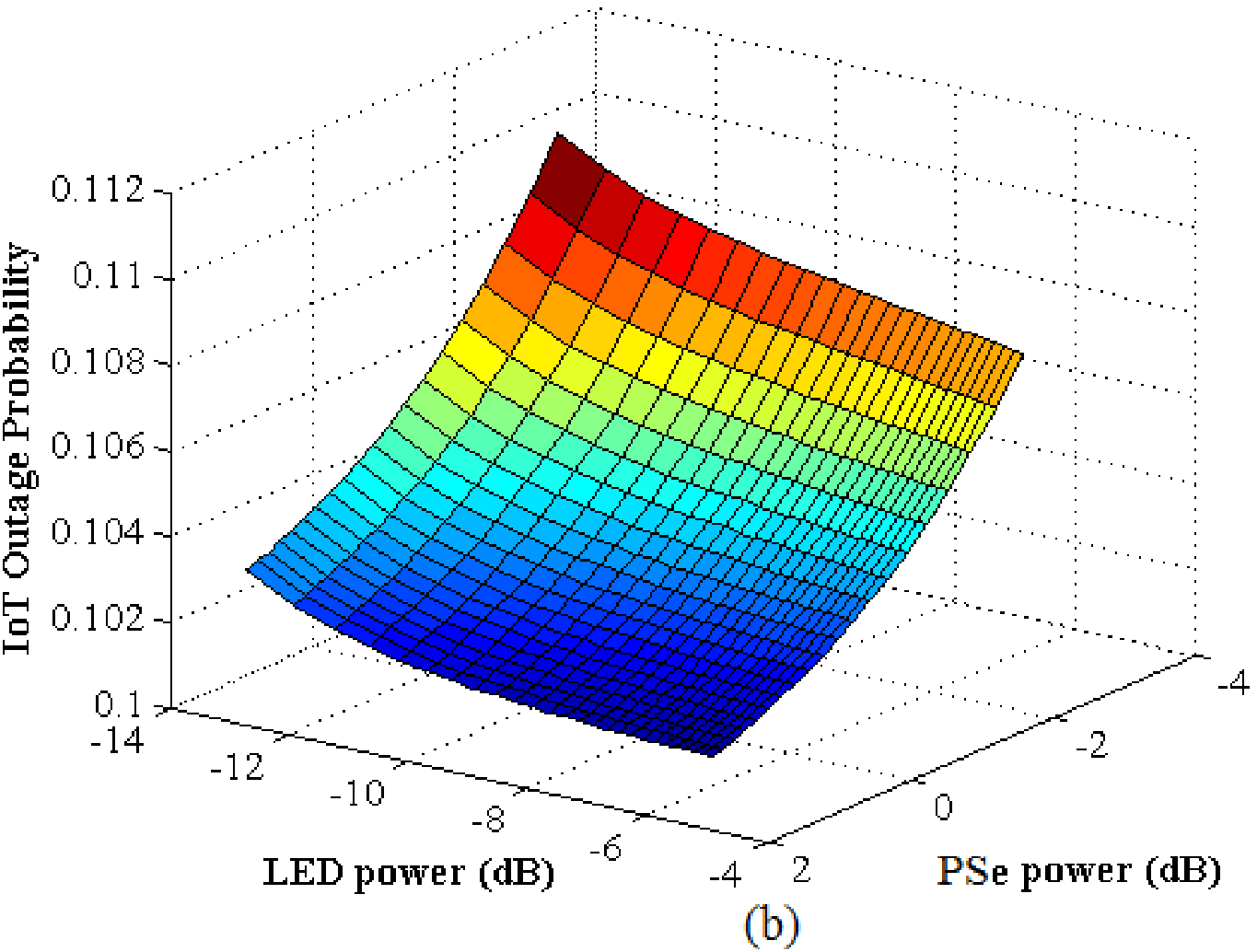}
         %\caption{\textbf{for different relaying techniques using N$_{tb}$=m$_k$=2}}
              \end{subfigure}
                \hfill
     \begin{subfigure}[b]{0.33\textwidth}
         \centering
         \includegraphics[width=\textwidth]{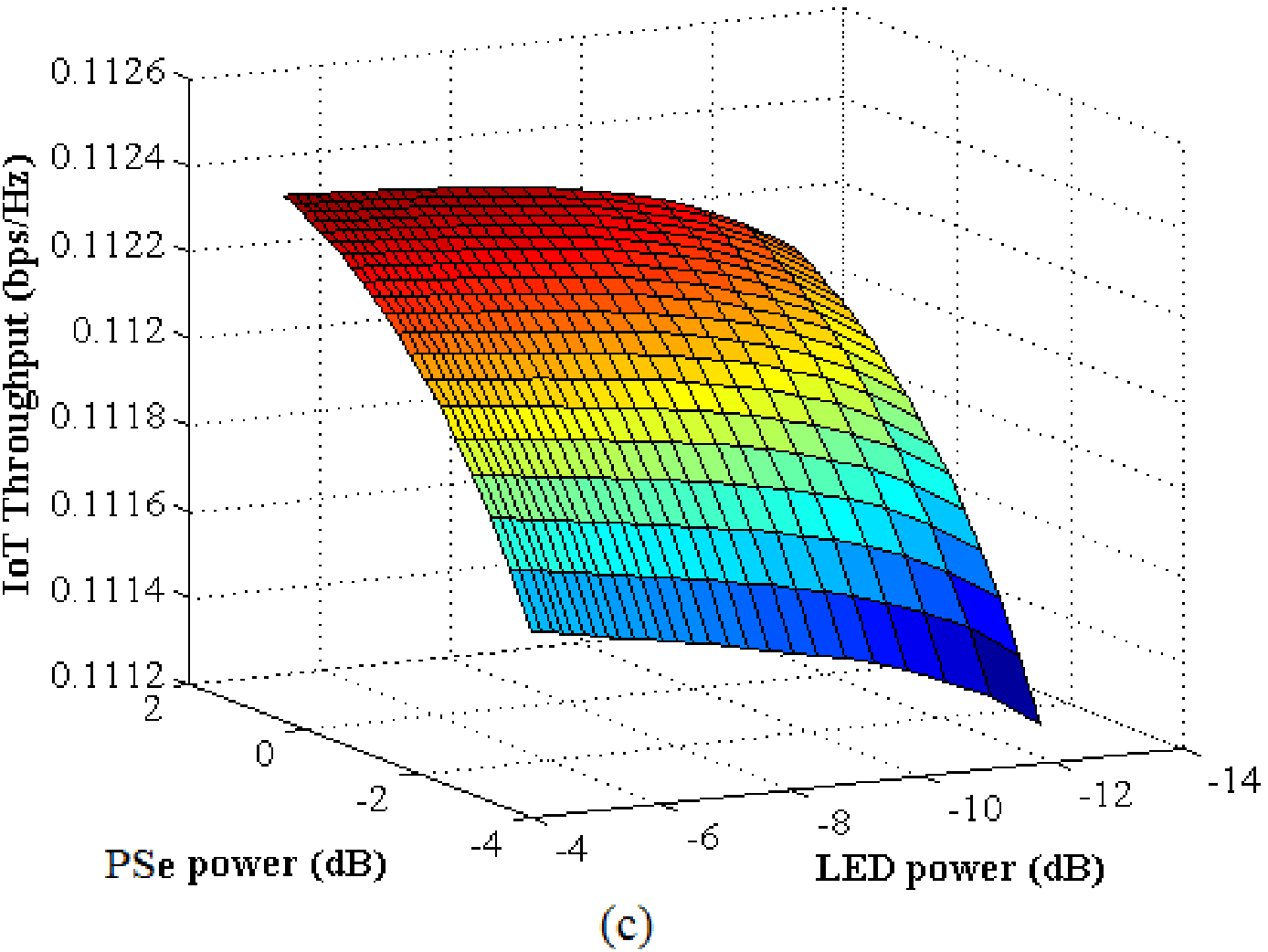}
         %\caption{\textbf{for different relaying techniques using N$_{tb}$=m$_k$=2}}
              \end{subfigure}
              \caption{{Outage probability vs PSe power vs LED power, (a) LU network, (b) IoT network, (c) IoT Throughput vs PSe power vs LED power}}
     \hfill
    \end{figure*}  
      \begin{figure*}[h]
     %\centering
     \begin{subfigure}[b]{0.325\textwidth}
         \centering
         \includegraphics[width=\textwidth]{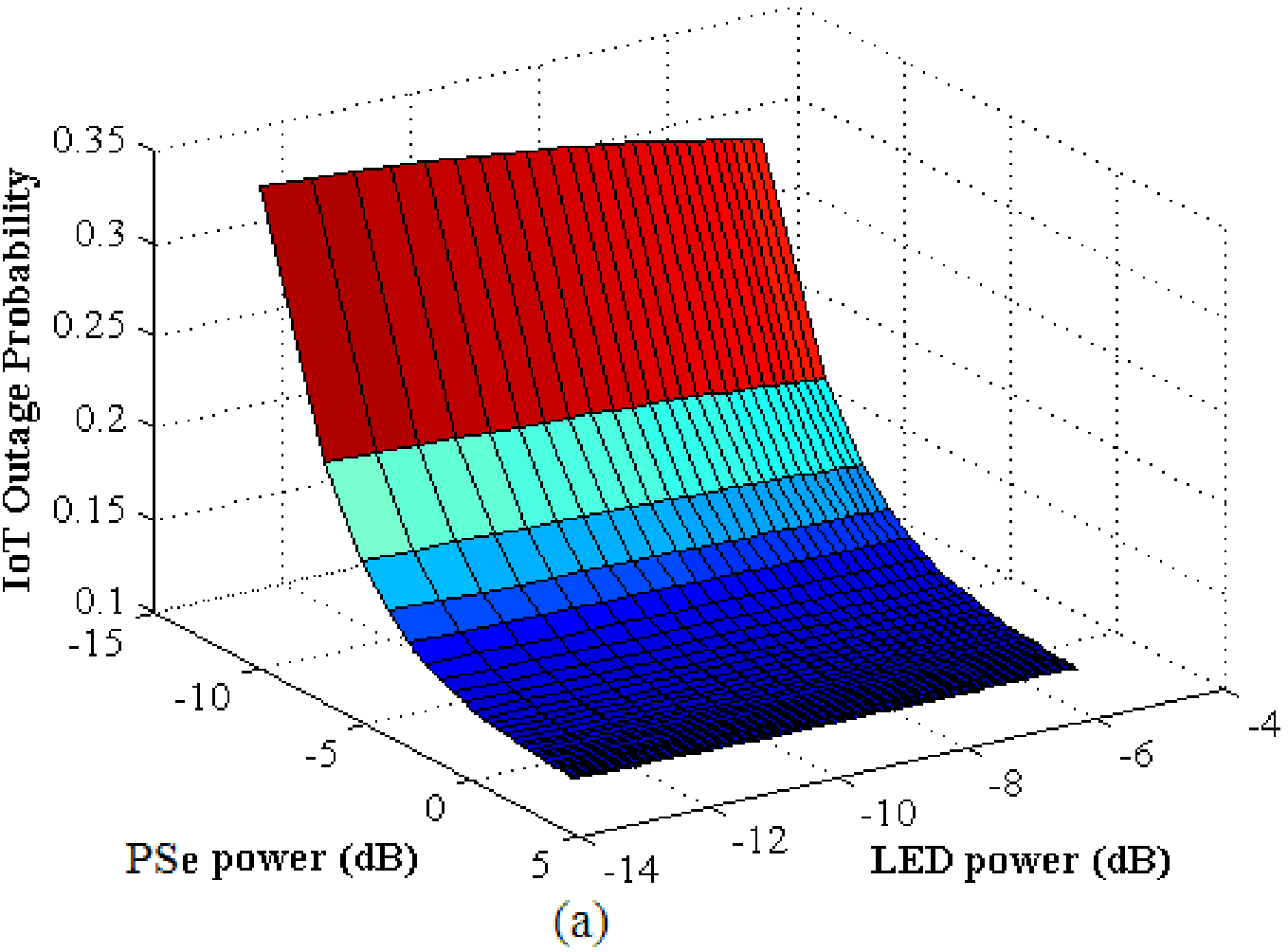}
        % \caption{\textbf{Outage probability vs $\lambda_1$=$\lambda_2$=$\lambda$ for different values of N$_{tb}$ and m$_k$=1 using hybrid relaying,}}
          \end{subfigure}
     \hfill
     \begin{subfigure}[b]{0.33\textwidth}
         \centering
         \includegraphics[width=\textwidth]{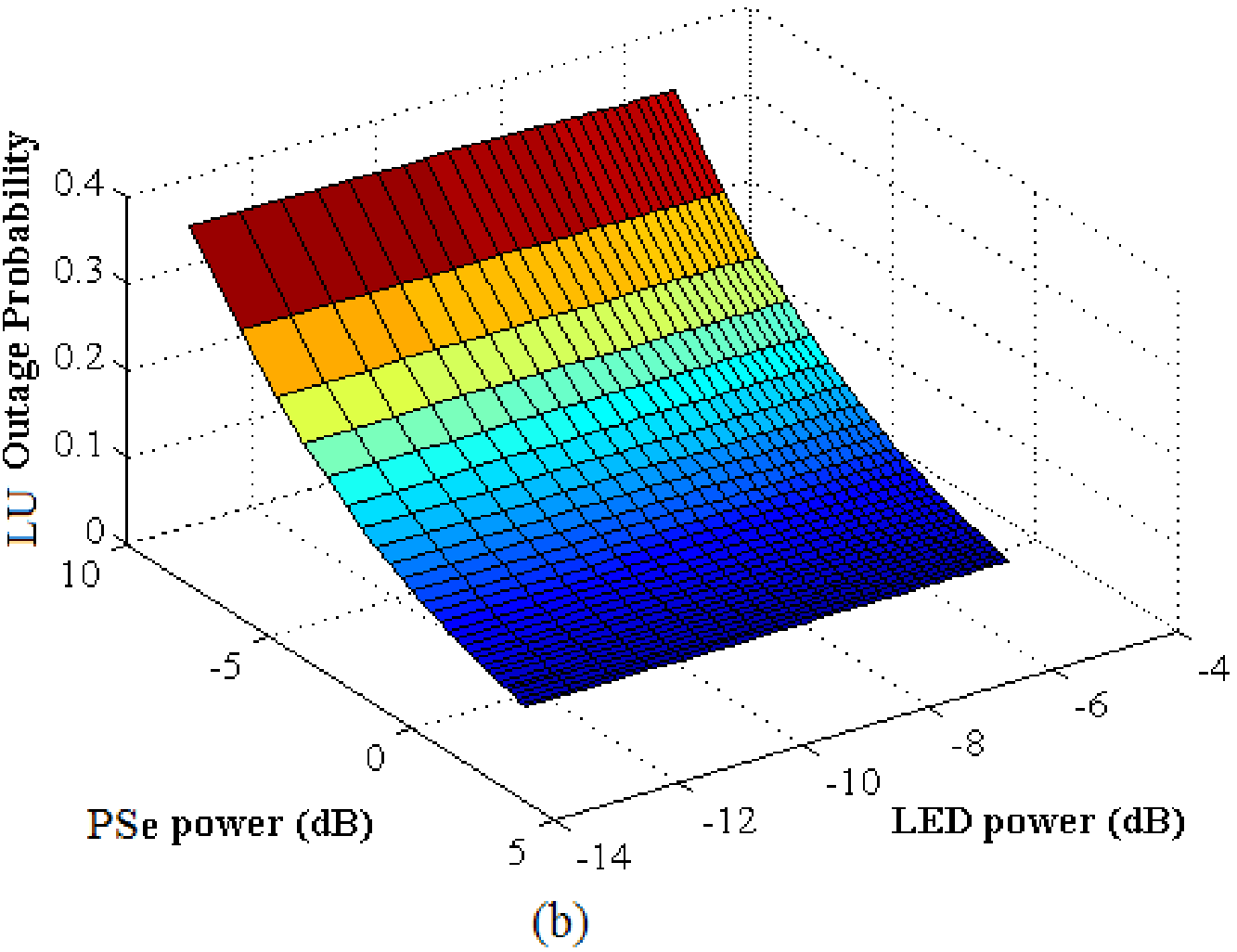}
         %\caption{\textbf{for different relaying techniques using N$_{tb}$=m$_k$=2}}
              \end{subfigure}
                \hfill
     \begin{subfigure}[b]{0.33\textwidth}
         \centering
         \includegraphics[width=\textwidth]{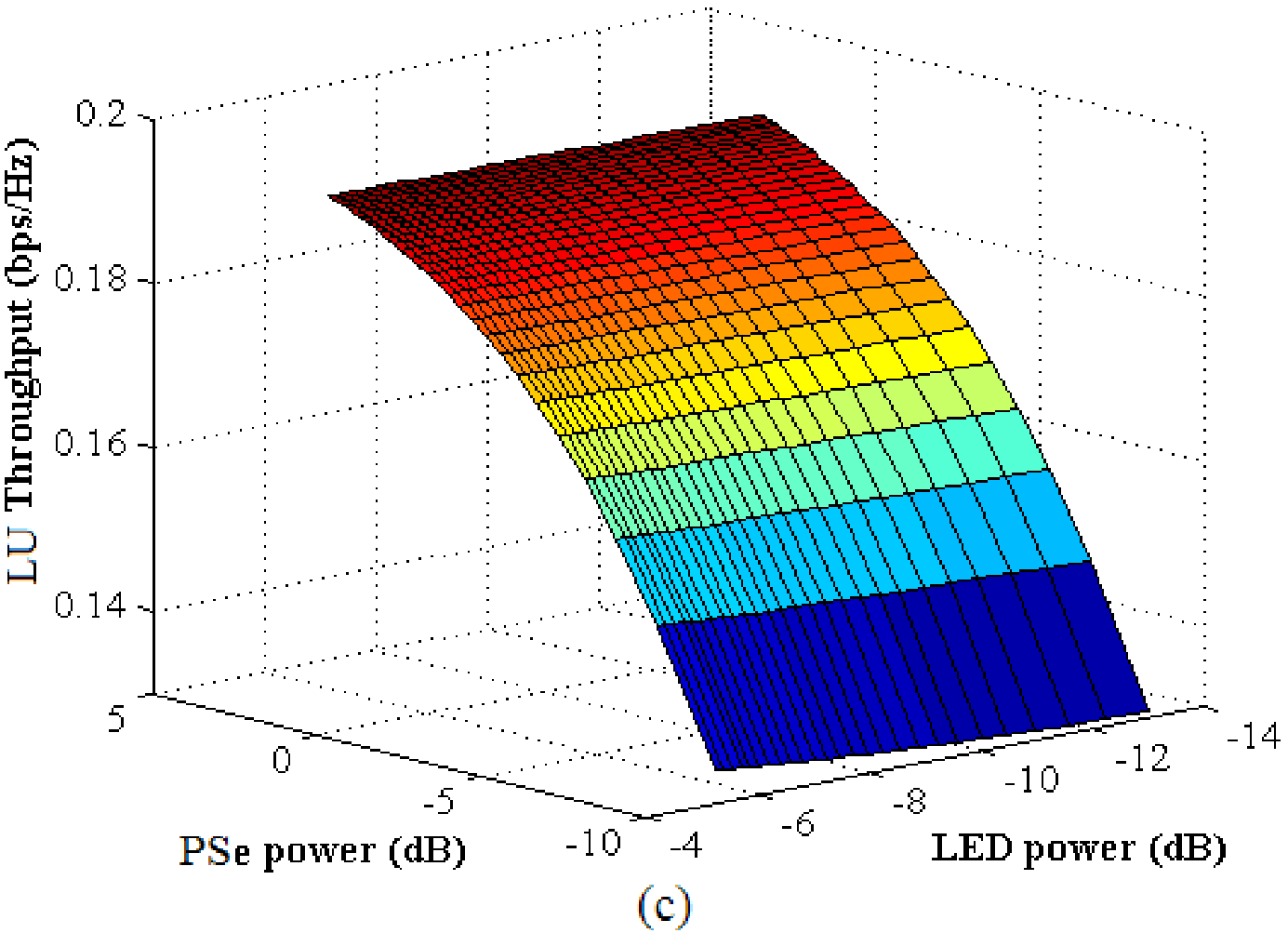}
         %\caption{\textbf{for different relaying techniques using N$_{tb}$=m$_k$=2}}
              \end{subfigure}
              \caption{{Outage probability vs PSe power vs LED power, (a) IoT network, (b) LU network, (c) LU Throughput vs PSe power vs LED power}}
     \hfill
    \end{figure*}   
 
 Fig. 4 depicts both LU and IoT outage performances with the variation of $A_p$ for different values of $\tau$. As achievable rate of information transmission from VT to AA is increased with the increase in $A_p$, therefore, both the outage performances are improved with increasing $A_p$. The worst outage performance is found at $\tau=0.5$ and the variation of IoT outage is insignificant as compared to LU outage for different values of $\tau$. LU outage performance is found to be the best at $\tau=0.1$. 
 
   Fig. 5.a, 5.b and 5.c show the LU network outage performance, IoT network outage and throughput performances, respectively with respect to the variation of both $\rho$ and $\tau$ for $N_L=20$, $P_{L}=200$ mW, $P_{bt}=1$ W, $R_{c1}=1.4$ bps, ${R_{c2}}=0.8$ bps, ${R_{c}}=0.4$ bps and $R_i=0.25$ bps. The nature of the plot of Fig. 5.a is exactly convex in nature. When $\tau$ is very less for fixed value of $\rho$, the achievable rate of VT$\rightarrow$AA is very poor. The outage performance is gradually improved with the increasing value of $\tau$ and it attains the minimum value for an optimum value of $\tau$. If $\tau$ is increased beyond 0.1, then the performance is degraded further. From Fig. 5.b, it is found that the performance of IoT outage is also improved with the increasing value of $\tau$ and it is also observed from the set of results that the performance is degraded when the $\tau$ is found to be very high. Even if the transmission power of PSe is very high, the performance of IoT outage is also degraded with the increasing value of target rate threshold.
  
   As shown in Fig. 5.a, the LU outage performance is found to be worse at $\rho \rightarrow 0$ and $\rho \rightarrow 1$ for some fixed $\tau$. Gradually, the outage performance is improved with the increase in $\rho$ and it reaches the minimum value for an optimum value of $\rho$. Further enhancement of $\rho$ results in the deterioration of LU performance in terms of outage. The reason behind this graphical plot can be demonstrated as follows. Initially, when $\rho$ is very small, the harvesting energy {{(following SWIPT protocol in Phase-2)}} is insufficient to transmit information from AA. The harvesting energy is improved with an increase in $\rho$. After reaching the minimum value of outage, if $\rho$ is increased again then the data transmission from PSe to AA is failed to meet the target rate of data transmission. Consequently, the outage performance is degraded. Over the certain range of $\rho$ ($\sim$ 0.58-0.99) and $\tau$ ($\sim$ 0.09-0.34), the outage probability of the LU network is found to be less than or equal to 10$\%$ outage threshold. This specific range is used to investigate outage probability and throughput of IoT network. As shown in Fig. 5.b and Fig. 5.c, the minimum IoT outage (9.56$\%$) and the maximum IoT throughput (0.113 bps/Hz) are found at $\tau=0.09$ and $\rho=0.98$, respectively. Similar plot is also obtained in Fig. 6.b to minimize the LU outage probability maintaining 20$\%$ IoT outage threshold constraint (in Fig. 6.a). Over the range of $\rho$ ($\sim$ 0.44-0.96) and $\tau$ ($\sim$ 0.09-0.69), IoT network  outage is found to be less than or equal to 20$\%$ outage threshold. As shown in Fig. 6.b and Fig. 6.c, the minimum LU outage (5$\%$) and the maximum LU throughput (0.19 bps/Hz) are found at $\tau=0.11$ and $\rho=0.96$.

Fig. 7.a, 7.b and 7.c illustrate the LU network outage performance, IoT network outage and throughput performances, respectively with respect to the variation of both VT LED power and PSe power for $N_L=25$, $R_{c1}=1.4$ bps, ${R_{c2}}=0.8$ bps, ${R_{c}}=0.4$ bps and $R_i=0.25$ bps. Outage plot is exactly intuitive in nature and the performances are improved with the increase in both LED power and PSe power. The outage performance is also reflected in the performance of IoT throughput. LU outage is found to be less than or equal to 10$\%$ outage threshold for $P_{L} \geq$ 50 mW and $P_m \geq$ 0.55 W. IoT outage and throughput are investigated over this region. As shown in Fig. 7.b and Fig. 7.c, the minimum IoT outage is found as 10$\%$ and the maximum IoT throughput is found as 0.1123 bps/Hz, respectively. Similar result is also obtained in Fig. 8.b to minimize LU outage maintaining 20$\%$ IoT outage threshold constraint (in Fig. 8.a).  
IoT outage is found to be less than or equal to 20$\%$ outage threshold for $P_{L} \geq$ 50 mW and $P_m \geq$ 0.1 W. As shown in Fig. 8.b and Fig. 8.c, the minimum LU outage is found as 4$\%$ and the maximum LU throughput is found as 0.192 bps/Hz, respectively.

 \section{Conclusion}
This work shows the efficacy of the hybrid VLC and RF system to enhance data rate with better reliability. Closed form outage expressions of both two-way LU and two-way IoT communications are derived for overlay mode of spectrum sharing technique of CCRN using DF relaying scheme. Exhaustive searching technique is applied to solve constrained outage minimization problems of both LU and IoT communications. Simulation results are used to validate the mathematical analysis of the proposed system. From the numerical results, it is found that optimum outage performance is possible to achieve by allocating more time for EH from RF source compared to VLC source.

\bibliographystyle{IEEEtran}
\bibliography{IEEEabrv,library}

\begin{IEEEbiography}[{\includegraphics[width=1in,height=1.25in,clip,keepaspectratio]{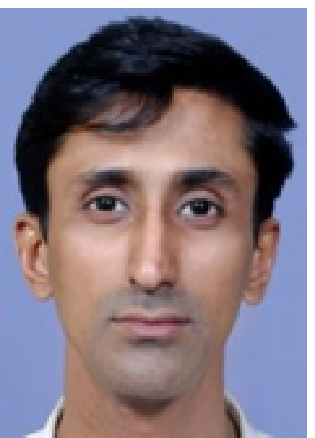}}]{Sutanu Ghosh}
(Member, IEEE) received his Ph.D. degree from Indian Institute of Engineering Science and Technology, India in 2021 and the M.Tech. degree from Jadavpur University, India in 2009. He is currently an Assistant Professor with the Department of Electronics and Communication Engineering, Institute of Engineering and Management, India. Currently, he is also working as a Visiting Scholar with the Computer, Electrical and Mathematical Science and Engineering Division, King Abdullah University of Science and Technology (KAUST), Thuwal, Saudi Arabia. His research interests include radio frequency energy harvesting, visible light communication, Internet-of-Things, cognitive radio networks.
\end{IEEEbiography}

\begin{IEEEbiography}[{\includegraphics[width=1in,height=1.25in,clip,keepaspectratio]{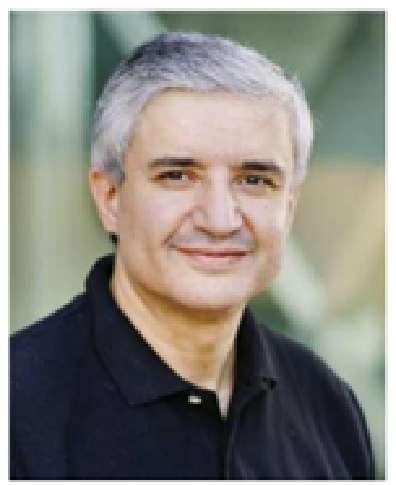}}]{Mohamed-Slim Alouini}
(Fellow, IEEE) was born in Tunis, Tunisia. He received the Ph.D. degree in electrical engineering from the California Institute of Technology, Pasadena, CA, USA, in 1998. 

He served as a Faculty Member with the University of Minnesota, Minneapolis, MN, USA, then in the Texas A{\&}M University at Qatar, Education City, Doha, Qatar, before joining King Abdullah University of Science and Technology, Thuwal, Saudi Arabia, as a Professor of Electrical Engineering in 2009. His current research interests include modeling, design, and performance analysis of wireless communication systems.
\end{IEEEbiography}

\end{document}